%% file: main.tex
\RequirePackage{fix-cm}
\documentclass[smallextended]{svjour3}       %
\usepackage{cite}
\usepackage{amsmath,amsfonts}
\usepackage[linesnumbered,ruled,vlined]{algorithm2e}
\usepackage{graphicx}
\usepackage{textcomp}
\usepackage{xcolor}
\usepackage{amssymb}
\usepackage{booktabs}
\usepackage{makecell}
\usepackage{listings}
\usepackage{multirow}
\usepackage{hyperref}
\usepackage{framed}
\usepackage{soul}
\usepackage[utf8]{inputenc}
\usepackage[T1]{fontenc}
\usepackage{microtype}
\usepackage{rotating}
\usepackage{paralist}
\usepackage{tabularx}
\usepackage{multicol}
\usepackage{pbox}
\usepackage{enumitem}	
\usepackage{colortbl}
\usepackage{pifont}
\usepackage{xspace}
\usepackage{url}
\usepackage{tikz}
\usepackage{fontawesome}
\usepackage{lscape}
\usepackage{color}
\usepackage{anyfontsize}
\usepackage{comment}
\usepackage{soul}
\usepackage{gensymb}
\usepackage{multibib}
\usepackage{balance}
\usepackage{footmisc}
\usepackage{xspace}
\usepackage[most]{tcolorbox}
 \usepackage{multirow}
 \usepackage{graphicx}
 \usepackage{booktabs}

\usepackage{lineno}

\usepackage[caption=false, font=footnotesize, position=top]{subfig}

\def\BibTeX{{\rm B\kern-.05em{\sc i\kern-.025em b}\kern-.08em
    T\kern-.1667em\lower.7ex\hbox{E}\kern-.125emX}}
    
\makeatletter
\def\lst@makecaption{%
  \def\@captype{table}%
  \@makecaption
}
\makeatother

\DeclareMathOperator*{\argmax}{arg\,max}
\DeclareMathOperator*{\argmin}{arg\,min}

\newboolean{showcomments}
\setboolean{showcomments}{true}         
\ifthenelse{\boolean{showcomments}}
  {\newcommand{\nb}[2]{
  \fbox{\bfseries\sffamily\scriptsize#1}
     {\sf\small$\blacktriangleright$\textit{\textcolor{red}{#2}}$\blacktriangleleft$}
   }
  }
  {\newcommand{\nb}[2]{}
   
  }

\newcommand{\gptf}{\textsc{GPT-4}\xspace}
\newcommand{\gptft}{\textsc{GPT-4T}\xspace}
\newcommand{\llama}{\textsc{Llama}\xspace}
\newcommand{\mixtral}{\textsc{Mixtral}\xspace}
\newcommand{\opus}{\textsc{Opus}\xspace}
\newcommand{\haiku}{\textsc{Haiku}\xspace}
\newcommand{\sonnet}{\textsc{Sonnet}\xspace}
\newcommand{\gptt}{\textsc{GPT-3.5T}\xspace}
\newcommand{\palm}{\textsc{PaLM}\xspace}

\newcommand\new[1]{{\color{black}#1}}

\newcommand{\COMMENT}[1]{}

\begin{document}

\title{Evaluating and Improving the Robustness of Security Attack Detectors Generated by LLMs}

\author{Samuele Pasini         \and
        Jinhan Kim  \and
        Tommaso Aiello \and
        Roc\'io Cabrera Lozoya \and
        Antonino Sabetta \and
        Paolo Tonella
}

\institute{Samuele Pasini \at
              Universit\`a della Svizzera italiana,
              Switzerland\\
              \email{samuele.pasini@usi.ch} 
           \and
           Jinhan Kim \at
              Universit\`a della Svizzera italiana,
              Switzerland\\
              \email{jinhan.kim@usi.ch} 
           \and
           Tommaso Aiello \at
              SAP Labs France,
              France\\
              \email{tommaso.aiello@sap.com} 
           \and
           Rocio Cabrera Lozoya \at
              SAP Labs France,
              France\\
              \email{rocio.cabrera.lozoya@sap.com} 
           \and
           Antonino Sabetta \at
              SAP Labs France,
              France\\
              \email{antonino.sabetta@sap.com}
           \and
           Paolo Tonella \at
              Universit\`a della Svizzera italiana,
              Switzerland\\
              \email{paolo.tonella@usi.ch} 
}

\date{Received: date / Accepted: date}

\maketitle

\begin{abstract}

Large Language Models (LLMs) are increasingly used in software development to generate functions, such as \textit{attack detectors}, that implement security requirements. 
\new{Ensuring the LLMs have enough knowledge to address specific security requirements, such as information about existing attacks, is a key challenge.}
For this, we propose an approach integrating Retrieval Augmented Generation (RAG) and Self-Ranking into the LLM pipeline. RAG enhances the robustness of the output by incorporating external knowledge sources, while the Self-Ranking technique, inspired by the concept of Self-Consistency, generates multiple reasoning paths and creates ranks to select the most robust detector. 
Our extensive empirical study targets code generated by LLMs to detect two prevalent injection attacks in web security: Cross-Site Scripting (XSS) and SQL injection (SQLi).  Results show a significant improvement in detection performance while employing RAG and Self-Ranking, with an increase of up to 71\%pt\footnote{\textit{Percentage points} are the standard unit of measure for differences between percentages.} \new{(on average 37\%pt)} and \new{ up to 43\%pt (on average 6\%pt)} in the F2-Score for XSS and SQLi detection, respectively. 

\keywords{Large Language Models \and Code Security \and Attack Detection \and Retrieval Augmented Generation}

\end{abstract}

\maketitle

\input{sections/introduction}

\input{sections/background}

\input{sections/approach}
\input{sections/empirical_study}

\input{sections/results}

\input{sections/discussion}

\input{sections/related_work}
\input{sections/threats}

\input{sections/conclusion}

\section*{Compliance with Ethical Standards}
\label{interest}
The authors declare that they do not have any known relationship or competing interests that could have influenced this paper.
The authors declare that the Human involved in the study for the assessment of the quality of the Generated Synthetic Dataset signed an informed consent approved by the Ethical Committee of Università della Svizzera Italiana.
The authors declare that their research for the current work did not involve Animals.

\section*{Data Availability}
\label{sec:avail}
The implementations, source code, data, and experimental results are publicly available in a GitHub repository\footnote{\url{https://github.com/PasiniSamuele/Robust-Attack-Detectors-LLM}}.

\section*{Credits}
\textbf{Samuele Pasini:} Problem Analysis, Investigation, Data Curation, Empirical Study, Writing - Original Draft, Visualization. \textbf{Jinhan Kim:} Investigation, Empirical Study, Writing - Original Draft. \textbf{Tommaso Aiello:} Empirical Study, Writing - Review \& Editing. \textbf{Rocio Cabrera Lozoya:} Writing - Review \& Editing. \textbf{Antonino Sabetta:} Resources, Writing - Review \& Editing. \textbf{Paolo Tonella:} Supervision, Writing - Review \& Editing.

\begin{acknowledgements}
This work is funded by the European Union's Horizon Europe research and innovation programme under the project Sec4AI4Sec, grant agreement No 101120393.
\end{acknowledgements}

\bibliographystyle{IEEEtran}
\balance

\bibliography{bibliography}

\end{document}

%% file: sections/introduction.tex
\section{Introduction}
\label{sec:intro}

The advent of Large Language Models (LLMs) has transformed software development with their impressive capabilities of understanding natural language prompts and producing accurate code that implements the given prompt.
These models are the foundation of AI-coding assistants like GitHub Copilot~\cite{copilot} or Cursor AI~\cite{cursorai}. With them, developers now often start with LLM-generated code as a base for refinement and testing~\cite{ghblog, tabachnyk2022productivity}.
However, this new practice also introduces potential risks: LLMs can inadvertently introduce vulnerabilities into the generated code or struggle to effectively generate \textit{security functions} that precisely satisfy the associated security requirements~\cite{perry2023userstudy, khoury2023secure, tihanyi2023swsecurity, nair2023hwsecurecode}. This issue likely stems from insufficient scrutiny of training data, lack of task-specific knowledge, inadequate fine-tuning, or missing assessment of the output~\cite{he2023large}.

An important family of security functions consists of \textit{attack detectors}.
With the words `attack detectors' we refer to the functions that, taking into account an input, possibly representing a payload, analyze the input and return a boolean value representing the outcome of the analysis. More specifically, they return a positive value if an attempt for an attack is found inside of the input payload. These functions require expert knowledge about the attack vectors for effective protection against such attacks.\footnote{Throughout the paper, we use the terms `attack detectors', `security functions', and `functions' interchangeably when discussing our approach. This flexible terminology enables us to define and apply our methodology across a range of domains.} This paper addresses this critical issue by evaluating and improving the robustness of the LLM-generated attack detection function. %

While the evaluation of LLM-generated code has garnered attention, most existing benchmarks, like HumanEval~\cite{chen2021evaluating}, emphasize complex algorithm generation rather than code generation requiring domain/security-specific knowledge. In addition, previous work has largely focused on \textit{identifying} vulnerabilities in LLM-generated code~\cite{khoury2023secure, mousavi2024investigation}, but a systematic approach to improving the robustness of security functions such as attack detectors is missing. We hypothesize that \textit{enhancing} the robustness of the generated detectors requires more than mere prompting with a suitable query -- it necessitates the integration of relevant knowledge, as well as the adoption of a systematic approach to robustness self-assessment.

To this end, we adopt two components to the LLM pipeline: Retrieval Augmented Generation (RAG)~\cite{lewis2020retrieval} and Self-Ranking. RAG, a technique widely investigated in Natural Language Processing (NLP), enhances the quality of the output by incorporating external information. 
We use RAG to leverage existing knowledge sources that document known attacks to increase the robustness of the detectors generated by LLMs.
Additionally, LLMs may exhibit non-deterministic behaviour in generation~\cite{ouyang2023llm}, which results in multiple, diverse solutions for the same prompt. 
We propose Self-Ranking, building upon the concept of Self-Consistency~\cite{wang2023selfconsistency}: we take advantage of the multiple reasoning paths associated with LLM's non-determinism to rank the alternative outputs and select the best one. More concretely, the LLM can propose different implementations obtained querying the model multiple times, with Self-Ranking we propose a strategy to rank the different implementations and to keep only the best-performing ones. In turn, ranking is based on the creation of a synthetic dataset, also generated by LLMs, used to automatically assess the robustness of each candidate output.

In our empirical study, we target two well-known and prevalent attacks: XSS and SQLi~\cite{HYDARA2015170,lawal2016systematic}, which are ranked second and third in `2023 CWE Top 25 Most Dangerous Software Weaknesses'\footnote{\url{https://cwe.mitre.org/top25/archive/2023/2023_top25_list.html}} respectively. %
Our extensive empirical study involves the evaluation of nine different LLMs. Results indicate that integrating external knowledge with RAG improved detection performance up to 66\%pt (on average 17\%pt) on the F2-Score for XSS detection and up to 67\%pt (on average 7\%pt) for SQLi detection compared to relying solely on few-shot examples. F2-Score is a metric which gives double importance to recall than to precision, more details about this metrics and the reasons behind its choice will be discussed in Section~\ref{sec:eval_metrics}.  Additionally, employing Self-Ranking enhanced the LLM performance by up to 71\%pt (on average 37\%pt) on the F2-Score for XSS detection and up to 43\%pt (on average 6\%pt) for SQLi detection. 

State-of-the-art (SOTA) Machine Learning (ML) based attack detectors require a labeled training set of attacks, while our LLM-based approach is applicable even when no training set is available, provided a RAG source is accessible. In our empirical study, we considered also the scenario when a training set is available and SOTA methods can be used. In such a scenario,
our approach achieved a performance comparable to SOTA XSS and SQLi machine learning-based detection methods, with the remarkable advantage that LLMs are pre-trained, while specialized machine learning-based methods require dedicated model design and training: developers can obtain SOTA performance by just querying an LLM using our RAG and Self-Ranking augmented pipeline.
Another key advantage is that LLMs generate code that can be understood and manipulated by developers, while the decision making logic of a machine learning model is to a large extent opaque to developers.
We also demonstrated the \textit{transferability} capabilities of LLMs, which requires just configuration steps: after configuring the LLM (i.e., setting parameters such as the model checkpoint to use, the temperature, the number of few-shot examples)  for optimal performance on one attack detection task (e.g., XSS), the resulting configuration can be transferred to another task (e.g., SQLi), with 16\%pt and 10\% improvement when transferring the LLM parameters from one task to another, as compared to average cases without parameter transfer.
On the other hand, transferability between tasks for the SOTA detectors would require additional training steps, as well as a dataset specific to the target vulnerabilities.

The technical contributions of this paper are as follows: 
\begin{itemize}
    \item We introduce an approach that integrates RAG and Self-Ranking to evaluate and improve the robustness of LLM-generated attack detectors.
    
    \item We conduct extensive empirical experiments with nine LLMs and two attacks, demonstrating the usefulness of our approach. %

    \item We explore the transferability of optimal parameters between the two tasks, contributing to a generalizable approach to securing LLM-generated code.
\end{itemize}

%% file: sections/background.tex
\section{Background}
\label{sec:background}

In this section, we provide a background on the two foundational concepts for our approach: RAG and Self-Consistency.

\subsection{Retrieval-Augmented Generation (RAG)}

LLMs often struggle with very specific topics, tending to produce \textit{hallucinations}~\cite{zhang2023siren}, making it challenging to ensure factual correctness.
This happens because the knowledge of the LLM is stored in its parameters (parametric memory), and it is highly compressed.
Retrieval-Augmented Generation (RAG)~\cite{lewis2020retrieval} tries to address this issue by incorporating a non-parametric memory, such as a database, to enhance the output's quality in knowledge-intensive tasks. RAG's key idea is to utilize external knowledge bases to fetch relevant information based on semantic similarity with the prompt, thereby reducing hallucinations. This approach has gained traction, making RAG essential in developing advanced chatbots \cite{gao2023retrieval}. 
The usage of RAG is also extremely useful to connect private data, not present in the initial training data, to an LLM without requiring any fine-tuning procedure.

A RAG application typically consists of two main steps. The first step is \textit{indexing}, a pipeline for ingesting data from one or more sources and creating an index. This process usually occurs offline for efficiency. During indexing, the data sources are broken into smaller chunks, since large chunks are harder to search over and would not fit into the model’s finite context window.
The result of this step is vector representations of the original knowledge source, obtained using an embedding model and stored in a vector database for efficient retrieval.
The second step is  \textit{retrieval}: when a user submits a prompt, RAG uses the same embedding model from the indexing phase to convert the prompt into a vector representation. It then calculates similarity scores between the prompt vector and the vectors of chunks in the indexed corpus. Then, RAG selects and retrieves the top $k$ chunks with the highest similarity to the prompt. These relevant chunks are then used to expand the context of the original prompt, providing additional information to the LLM.

\subsection{Self-Consistency}
LLMs are limited in their reasoning capabilities and this cannot be resolved merely by scaling them up~\cite{rae2021scaling, srivastava2022beyond}.
One of the most relevant approaches to overcome this limitation is Chain-of-Thought prompting (CoT)~\cite{wei2022chain}, which prompts LLMs to produce a sequence of short sentences replicating a human's reasoning process to solve a task.

The idea of CoT was further extended into Self-Consistency~\cite{wang2023selfconsistency},
based on the observation that complex reasoning tasks often allow multiple reasoning paths to reach a correct answer~\cite{stanovich2012distinction}. The LLM is first prompted with CoT. Then, instead of greedily decoding the optimal reasoning path, a new ``sample-and-marginalize'' decoding strategy generates a diverse set of reasoning paths, each of which potentially leads to a different answer. The final answer is determined by marginalization, i.e., by summing up probabilities over these paths to find the most consistent response. Importantly, Self-Consistency differs from a typical ensemble approach, as it operates on a single LLM (also called a ``self-ensemble''), rather than aggregating outputs from multiple LLMs~\cite{wang2023selfconsistency}.

%% file: sections/approach.tex
\begin{figure*}[!pht]
  \centering
  \includegraphics[width=1.5\linewidth ,angle=270,origin=c]{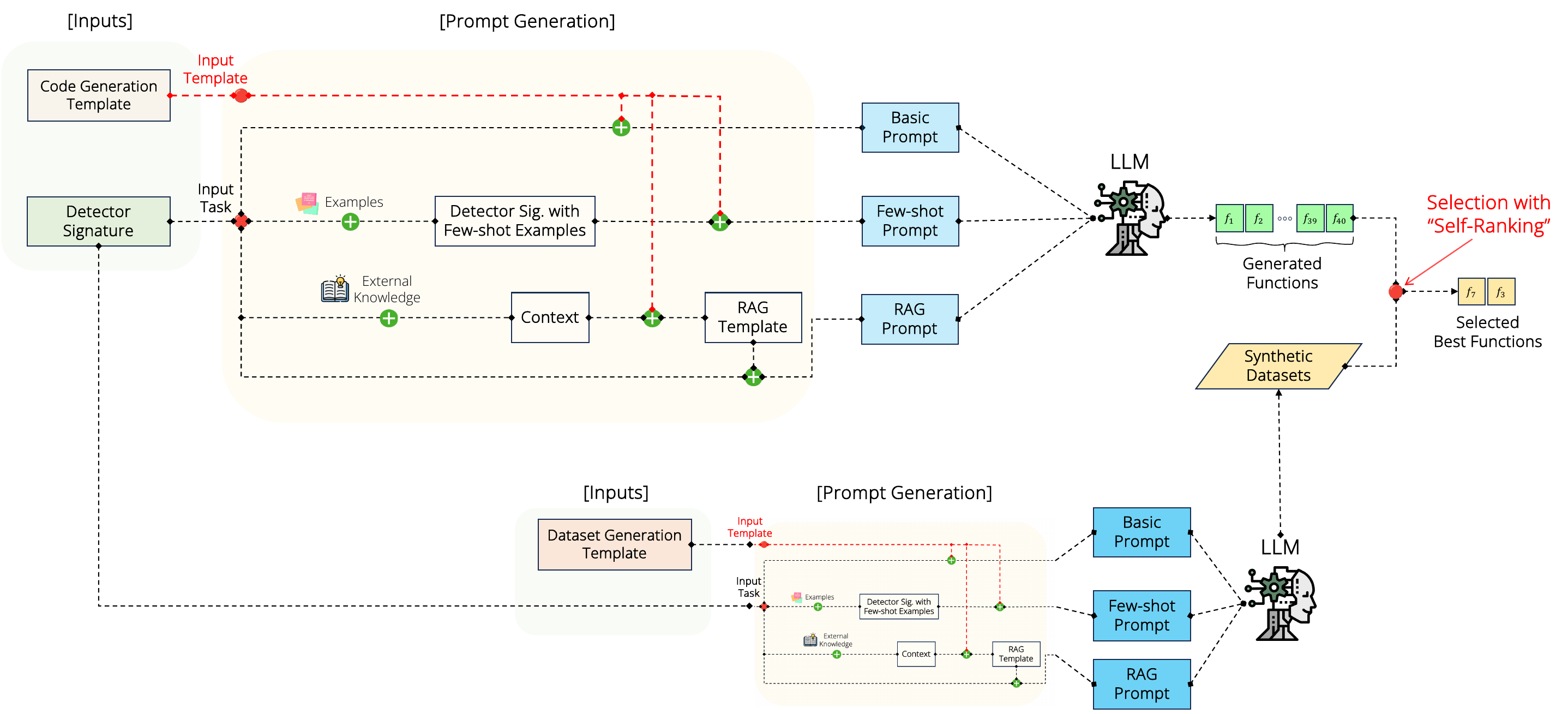}
  \caption{%
  Overview of our approach. There are two large parts, one for code generation (left) and the other for dataset generation (bottom-right). Each of the two parts takes a `Code Generation/Dataset Generation Template' and a `Detector Signature' as input (respectively matching `Input Template' and `Input Task') and generates one of four types of prompts: \textit{Basic}, \textit{Few-shot}, \textit{RAG}, or \textit{RAG Few-shot} (for simplicity, the latter is omitted). The prompt is fed into an LLM to generate candidate functions (top pipeline) or a synthetic dataset (bottom pipeline). We then use a Self-Ranking mechanism to evaluate and rank the generated functions based on their performance on the synthetic dataset.}\label{fig:approach}
\end{figure*}

\section{Approach}
\label{sec:approach}

In this section, we present our approach to assess and improve the robustness of LLM-generated attack detectors, achieved through a combination of LLM prompt design, RAG, and Self-Ranking, a novel technique built upon Self-Consistency.
Figure~\ref{fig:approach} illustrates an overview of our approach. A \textit{Prompt} (shown in the middle within a blue box) is constructed and fed into the LLM to generate functions, in our case attack detectors (top pipeline) or synthetic datasets, used to self-assess the generated functions (bottom pipeline). We present four types of Prompts, \textit{Basic}, \textit{Few-shot}, \textit{RAG}, and \textit{RAG Few-shot}\footnote{For the sake of simplicity, the \textit{RAG Few-shot} Prompt is not shown in Figure~\ref{fig:approach}. It can be easily obtained by combining the construction of \textit{Few-shot} and \textit{RAG} Prompts.}, which vary in their use of RAG and Few-shot examples. The synthetic dataset is then used to evaluate and select the generated functions by means of the Self-Ranking mechanism (described in the following sections). In the following subsections, we detail the construction of each type of Prompt and explain how RAG and Self-Ranking are employed.

\subsection{Basic Prompting}

We provide the LLM with a Prompt that consists of two parts: \textit{Template} and \textit{Task} (in Figure~\ref{fig:approach} (left), these are illustrated as `Code/Dataset Generation Template' and `Detector Signature', respectively). The Template contains a general instruction that specifies the desired output. To generate Python functions, such as attack detectors, a `Code Generation Template' is used, which instructs the LLM to produce Python code as shown in Listing~\ref{lst:code_template}.
\begin{lstlisting}[label={lst:code_template}, caption={Code Generation Template}]
The user will provide the initial part of a python function (function name, parameters with types, return type and a comment describing the purpose of the function) with some optional example. 
You are a coding assistant that writes some python code to create the user's function. Be sure that the code is syntactically correct, it is a callable function (containing the initial part provided by the user) and that it returns the correct type. 
The length of the code should be short, it should be readable, and without redundant checks on the parameters. 
Return only python code in Markdown format, e.g.:
```python
....
```
\end{lstlisting}\leavevmode

The Task contains the \textit{signature} of the function (e.g., attack detector) followed by a short comment describing its purpose. %
Two examples of tasks we consider in our empirical study, XSS and SQLi, are shown in 
Listing~\ref{lst:xss_task} \& \ref{lst:sqli_task}, respectively.

\lstset{escapeinside={(*@}{@*)}}
\begin{lstlisting}[language=Python, label={lst:xss_task}, caption={Task for XSS detection (Detector Signature)}]
def detect_xss(http_get_request: str)->bool: 
""" Check if in the given http_get_request there is an XSS exploit, considering also the possible evasions that an attacker can perform.
"""
\end{lstlisting}\leavevmode

\lstset{escapeinside={(*@}{@*)}}
\begin{lstlisting}[language=Python, label={lst:sqli_task}, caption={Task for SQLi detection (Detector Signature)}]
def detect_sqli(query: str)->bool: 
""" Check if the given SQL query contains some statement injected by an attacker to perform SQL injection. Be sure that in the patterns used for the detection, no pattern related to normal SQL code (like standard SELECT ... FROM ... WHERE ..., or standard INSERT ... INTO ...) is considered, since it is not an attack."""
\end{lstlisting}\leavevmode

The \textit{Basic Prompt} is the result of merging  Template and Task, serving as the input to the LLM. For example, we can combine Listing~\ref{lst:code_template} and Listing~\ref{lst:xss_task} to form a Basic Prompt for the generation of an XSS detector.

\subsection{Few-shot Examples}

Few-shot examples have been explored in benchmarks such as HumanEval~\cite{chen2021evaluating}, where they are appended to the function definition to illustrate the expected input-output relationship. In our approach, we extend this concept by including malicious and benign examples, creating what we call the \textit{Few-shot Prompt}. An illustration of this is shown in Listing~\ref{lst:xss_task_examples}, where we use Few-shot examples for XSS detection. Note that while an arbitrary number of examples can be appended, there is no guarantee that a higher quantity will yield better results.

\lstset{escapeinside={(*@}{@*)}}
\begin{lstlisting}[language=Python, label={lst:xss_task_examples}, caption={Example task for XSS detection with few-shot examples}]
def detect_xss(http_get_request: str)->bool: 
""" Check if in the given http_get_request there is an XSS exploit, considering also the possible evasions that an attacker can perform.
>>> detect_xss('http://anonymised-site/moderate-social-media&t=1396533765893&n=1129109&k=mainentity')
False
>>> detect_xss('http://anonymised-site/guestbook/index.php?lang=&quot;&gt;&lt;script&gt;alert(document.cookie)&lt;/script&gt;...')
True
"""
\end{lstlisting}\leavevmode

\subsection{RAG}

We followed the general RAG methodology, starting from the indexing phase.  Once the knowledge sources about attacks are stored in vector representation\footnote{The selection of appropriate data sources and their storage is discussed in Section~\ref{sec:empirical_study} as it is an experimental choice.}, they can be queried to retrieve the most relevant information chunks (referred to as  `External Knowledge' in Figure~\ref{fig:approach}). We utilize the Task (`Detector Signature' in Figure~\ref{fig:approach}) directly as a query to retrieve these chunks, which are then appended to the Template to obtain a \textit{RAG Prompt}, as illustrated in Listing~\ref{lst:code_template_rag}.
\new{We chose to append the retrieved chunks to the template.} %
\new{While external knowledge could be added to the function's comments, we found this approach to be less fluent and practical, especially given the size of the retrieved knowledge.}

Few-shot examples and RAG can be combined as they modify different structures: the Task and the Template, respectively. Appending Few-shot examples to the Task would allow RAG to extract more diverse and useful information, possibly enhancing the overall performance. We call this combined prompt as \textit{RAG Few-shot Prompt}.

\lstset{escapeinside={(*@}{@*)}}
\begin{lstlisting}[label={lst:code_template_rag}, caption={Template for code generation, with the relevant context selected via RAG (queried with XSS Detection Task) appended at the end. The first part of the template is shortened to avoid redundancy.}]
The user will provide the initial part of a python function 
...
...
Return only python code in Markdown format, e.g.:
```python
....
```
Use the following pieces of retrieved context to write a more complete function:
Context: Here's an XSS example that bets on the fact that the regex won't catch a matching pair of quotes but will rather find any quotes to terminate a parameter string improperly.
...
Example: <script> ... setTimeout(\\"writetitle()\\",$\_GET\[xss\]) ... </script>
Exploitation: /?xss=500); alert(document.cookie);//)
\end{lstlisting}\leavevmode

\subsection{Self-Ranking}

Self-Consistency does not apply directly to our task as proposed originally~\cite{wang2023selfconsistency}, because the generation of security attack detectors is not always easily decomposable into a chain of thoughts, representing partial and incremental steps leading to the solution. 
Moreover, reasoning on the generated candidates may not be feasible, as the set of all the generated functions may not fit within the LLM's context window. More importantly, our preliminary experiments suggest that LLMs struggle to assess the robustness of a set of candidate functions.
To explore this further, we applied the LLM-as-a-judge approach~\cite{zheng2023judging}, prompting \gptf to rate the robustness of several sampled detectors on a scale from 0 to 4. Over 95\% of the scores fell within the top two categories (3 and 4), indicating poor discrimination among candidates. To assess whether these scores aligned with actual performance, we conducted a Spearman Correlation test between the LLM-assigned ranks and the F2-scores of the same detectors on a validation set. The correlation was negligible. A subsequent linear regression analysis, using the LLM score as the independent variable and F2-score as the dependent variable, yielded a 95\% confidence interval ranging from $-0.09$ to $0.11$, reinforcing the conclusion that no meaningful relationship exists. These findings highlight the inadequacy of direct LLM-based evaluation for robustness scoring. Hence, we reuse only the general idea behind Self-Consistency, i.e., the LLM operating as a ``self-ensemble''. Specifically, the non-determinism of LLMs is leveraged to produce a set of candidates that are assessed by the LLM itself, which generates an assessment dataset instead of directly assessing its own output. 
We refer to our novel variant of Self-Consistency as \textit{Self-Ranking}: we ask the LLM to generate a synthetic dataset that simulates the presence of ground-truth data to select the best function. This synthetic dataset is used to evaluate and rank the generated functions. %

To generate the synthetic dataset, we leverage the modular structure of the Prompt by introducing a second Template (`Dataset Generation Template' in Figure~\ref{fig:approach}), as shown in Listing~\ref{lst:dataset_template}. This `Dataset Generation Template' can be combined with the two proposed Tasks to create synthetic datasets for various vulnerabilities. Additionally, Few-shot examples and RAG can be integrated into synthetic dataset generation.

\begin{lstlisting}[label={lst:dataset_template}, caption={Dataset Generation Template}]
The user will provide the initial part of the function (function name, parameters with types, return type and a comment describing the purpose of the function, with some optional example. 
You are a testing assistant that generates a dataset to test the function provided by the user.

\end{lstlisting}\leavevmode

Once the synthetic dataset is generated, it serves as a testing ground for selecting the best function. Specifically, when multiple functions are generated, Self-Ranking starts evaluating each function with the synthetic dataset, ordering them based on a chosen metric, e.g., F2-Score, and retaining only the top-performing subset, denoted as the \textit{top\_k} functions. %
In our study, the usage of Self-Ranking aims to select the subset of  $k$ most robust attack detectors. %
\new{Unlike the LLM-as-a-judge methodology, our Self-Ranking approach does not ask one LLM to directly judge another's output. Instead, we use synthetic data generated by an LLM to evaluate the performance of our detectors.}

%% file: sections/empirical_study.tex
\section{Empirical Study}
\label{sec:empirical_study}

In this section, we first present the two evaluation scenarios, which differ in the availability vs. unavailability of a training/validation dataset. We then formulate four research questions and describe our experimental settings, including the models used, the configurable parameters, and the datasets. Next, we outline the experimental procedure, including the generation and evaluation of functions, the generation of synthetic datasets, and the selection of \textit{top\_k} functions. Finally, we provide details on the implementation and the evaluation metrics.

\subsection{Scenarios}

\subsubsection{Scenario NTD (No Training Dataset)}
In this scenario, developers do not possess a training set that would allow them to train a machine learning-based SOTA detector. For the same reason, developers cannot extract a validation set from the training set, which would allow them to directly assess and rank the LLM-generated functions (without Self-Ranking). Also, they cannot decide if one LLM (checkpoint) is preferable over another one (assuming that they have multiple instances to select from),  find optimal LLM parameter configurations (e.g., temperature), and choose the optimal prompt among Basic/Few-shot/RAG/Rag Few-shot Prompts. Moreover, they cannot decide on the potential benefits offered by Self-Ranking. 

Therefore, in the NTD scenario, we consider the effectiveness of our approach in terms of \textit{average} performance on a test set composed of real-world attacks. We average the performance calculated across multiple choices of LLM instances,  temperatures, alternative function/dataset generation prompts, and inclusion/exclusion of Self-Ranking. The \textit{average} performance can be conditioned on each element of our approach, to determine its effect on the average performance when all other choices are unconstrained. %

\subsubsection{Scenario TDA (Training Dataset Available)} 

In this scenario, a task-specific labeled training set of attacks is available to developers. This addition would allow them to directly train ML models that perform the task assigned to the generated detectors, without actually generating any function at all. In fact, SOTA techniques for many security tasks, including XSS and SQLi detection, use ML models trained on a labeled dataset. 

Moreover, the availability of a training dataset allows the extraction of a validation set that can be used to analyze the performance of different LLM instances,  parameters, and elements of our approach, to choose the optimal configuration.  Developers may even choose between LLM-generated functions and  ML model, based on the respective performance on the validation set.

\subsection{Research Questions}
\label{sec:rqs}
In both NTD and TDA scenarios, we want to know if RAG and/or Self-Ranking are beneficial to the generation of robust attack detectors: 

\begin{itemize}
    \item \textbf{RQ1. RAG Benefit}: \textit{How helpful is RAG in generating better security attack detectors? How does it perform when combined with Few-shot examples?}
    \item \textbf{RQ2. Impact of Self-Ranking}: \textit{Is the selection of \textit{top\_k} functions via Self-Ranking an effective strategy to enhance the robustness of LLM-generated attack detectors?}
\end{itemize}

In the NTD scenario, we evaluate the average performance across configurations (LLM, temperature, number of few-shot examples), either with or without RAG/Self-Ranking. In the TDA scenario, we check whether the optimal configuration selected through the validation set includes RAG/Self-Ranking.

In the TDA scenario, the available training set can be used to train  a SOTA machine learning-based model. The best LLM-generated functions can be compared with SOTA techniques solving the same task:

\begin{itemize}
    \item \textbf{RQ3. Comparison with SOTA}: \textit{Do the detectors generated by LLMs, when assessed on an existing validation dataset, demonstrate comparable performance to state-of-the-art ML models trained specifically for the task?}
\end{itemize}

Finally, we are interested in the transferability from the TDA scenario on a task, to the NTD scenario on another task. We want to know if the optimal configuration of the LLM identified through the validation set available for a given task (e.g., XSS) provides good/optimal results when used to solve another security function generation task (e.g., SQLi).

\begin{itemize}
    \item \textbf{RQ4. Transferability}: 
    \textit{Can the optimal parameters for function and synthetic data generation in one task be transferred and applied to achieve effective results in the other task?}
\end{itemize}

\begin{table*}[h]
\caption{LLMs used in the experiments. Column C. Window  shows the size of the Context Window in tokens; Up To shows the last training update.}
\label{tab:models}
\resizebox{\textwidth}{!}{%
\begin{tabular}{r|r|r|r|r|r|r|r}
\toprule
\multicolumn{1}{c|}{Model Name} &
  \multicolumn{1}{c|}{Checkpoint Name} &
  \multicolumn{1}{c|}{Alias} &
  \multicolumn{1}{c|}{Provider} &
  \multicolumn{1}{c|}{N. parameters} &
  \multicolumn{1}{c|}{C. Window} &
  \multicolumn{1}{c|}{Up To} &
  \multicolumn{1}{c}{Pass@1}\\
  \midrule
GPT-3.5 Turbo     & \textit{gpt-3.5-turbo-0125}        & \gptt    & OpenAI       & N/A                                       & 16,385  %
& 2021 & 64.9 \\
GPT-4             & \textit{gpt-4-1106-preview}        & \gptf    & OpenAI       & N/A                                       & 128,000 %
& 2023  & 76.5   \\
GPT-4 Turbo       & \textit{gpt-4-0125-preview }       & \gptft   & OpenAI       & N/A                                       & 128,000 %
& 2023 & 87.1 \\
Claude 3 Haiku    & \textit{anthropic-claude-3-haiku}  & \haiku   & Anthropic    & Small Claude3     & 200,000 %
& N/A   & 75.9           \\
Claude 3 Sonnet   & \textit{anthropic-claude-3-sonne}t & \sonnet  & Anthropic    & Med. Claude3 & 200,000 %
& N/A     & 73.0         \\
Claude 3 Opus     & \textit{anthropic-claude-3-opus}   & \opus    & Anthropic    & Large Claude3  & 200,000 %
& N/A  & 84.9            \\
Llama3            & \textit{llama3-70b-instruct }      & \llama   & Meta         & 70 billions                              & 8,192   %
& N/A     & 81.7         \\
Mixtral 8x7b      & \textit{mixtral-8x7b-instruct-v01} & \mixtral & Mistral      & 12 billions                              & 32,000  %
& N/A      & 40.2        \\
PaLM 2 Chat Bison & \textit{gcp-chat-bison-001 }       & \palm    & Google & N/A                                       %
& 2,500  & 2023 & 37.6 \\
\bottomrule
\end{tabular}%
}
\end{table*}

\subsection{Experimental Settings}
\label{sec:exp_settings}

\subsubsection{Models}

We employ nine LLMs as listed in Table~\ref{tab:models}, which includes the aliases that are used throughout the remainder of this paper. At the time we started the experiments, we selected the best performing models available to us \new{through the SAP AI Core service, which allowed us access a broad range of state of the art LLMs. We further filtered the available models  based on their performance on the} HumanEval~\cite{chen2021evaluating}  benchmark. We also consider some of the weaker models to better highlight the differences of results in the considered task. Additionally, we report the Pass@1 scores from HumanEval as a proxy for their general reasoning capabilities: \gptft showed the best performance, followed closely by \opus, whereas \palm exhibited the lowest performance among the models we considered. 
While  Pass@1, which assesses the model's reasoning capabilities to generate complex algorithms, may serve as an indicator of its expected performance on our task, we do not expect a perfect correlation with our results, since our task is more focused on utilizing extensive knowledge rather than reasoning on a given problem.
\subsubsection{Configurable Parameters}

A \textit{model configuration} refers to the model and temperature used (e.g., \gptf with temperatures $0.5$ and  $1.0$). A \textit{prompt configuration} is determined by the number of Few-shot examples (with equal malicious and benign examples) and the choice to use RAG or not, to enrich the prompt. By combining model and prompt configurations, we obtain an overall Configuration \textit{Conf}, consisting of a tuple with a model, a temperature, the few-shot number, and the RAG usage choice. 
For instance,  $( \gptf, 0.5, 2, T)$ indicates the configuration where the model is \gptf, the temperature is set to 0.5, the number of Few-shot examples is 2, and RAG is used (T represents True for the RAG Usage Parameter).

Table~\ref{tab:confs} shows the domains of the configurations for the different parts of the empirical study.
Given the domain, some of its configurations may fail on a given task. Hence, they are excluded. By failure, we mean that the generation process did not produce a valid artifact.  Specifically, for Code Generation, failure occurs when more than 80\% of the generated functions raise exceptions upon execution. For Dataset Generation, a configuration fails if it cannot produce a valid CSV file meeting the specified requirements (e.g., number of samples and class distribution) within three hours.
For example, among the configurations for Code Generation (see Table~\ref{tab:confs}), when generating an XSS detector, all the configurations with models \sonnet, \palm, \llama, \mixtral, and temperature 0 fail and are hence discarded.
Similarly, for the SQLi detector, the aforementioned configurations, along with \palm, \sonnet, and \mixtral at temperature 0.5, fail and are also excluded.
\haiku and \gptt are completely discarded in code generation for the same reasons, but they are still among the employed models because they are used for synthetic dataset generation.

\begin{table}[h]
\caption{Domain of the configurations used for code and dataset generations}
\label{tab:confs}
\resizebox{\linewidth}{!}{%
\begin{tabular}{r|r|r|r|r}
\toprule
\multicolumn{1}{c|}{Conf. Domain} &
  \multicolumn{1}{c|}{Model} &
  \multicolumn{1}{c|}{Temp.} &
  \multicolumn{1}{c|}{NShot.} &
  \multicolumn{1}{c}{RAG.} \\
  \midrule
\makecell[c]{Code Generation} &
  \makecell[c]{\gptft, \gptf, \opus,\\ \sonnet, \palm,\\ \llama, \mixtral} &
  0.0, 0.5, 0.1 &
  0, 2, 6, 10 &
  True, False \\
  \midrule
\makecell[c]{Dataset Generation} &
  \makecell[c]{\gptft, \gptt, \opus,\\\sonnet, \haiku} &
  0.0, 0.5, 0.1 &
  0, 2, 6, 10 &
  True, False\\
  \bottomrule
\end{tabular}
}
\end{table}

Every time we generate code, we produce 40 functions by feeding the model 40 times with the same prompt. This strategy accounts for the non-determinism of LLMs~\cite{ouyang2023llm} and is exploited by Self-Ranking to select the \textit{top\_k} functions.

For synthetic dataset generation, we iteratively generate a dataset of 100 examples, with an equal split of 50 malicious and 50 benign cases. Specifically, we prompt the LLMs to generate datasets in this format, although we do not enforce any hard constraints to guarantee strict adherence to the desired class distribution. Because of that, we do not have any guarantee about the correctness of the labels of the samples produced by the LLMs. We use a timeout of 9,000 seconds to discard configurations that fail to fill the dataset with 100 examples in a reasonable amount of time. %
To account for non-determinism, we repeat the process 10 times and produce 10 distinct datasets.
The performance used to determine the ranking is calculated by averaging the performance on the 10 different produced datasets.
When Self-Ranking is used, we have to decide the number $k$ of top-ranked functions to consider. 
In our experiments, the values of $k$ for the selection of \textit{top\_k} functions are 1, 3, and 5.

Several RAG sources were analyzed with preliminary experiments to select the most suitable ones for our setup.
Considering only one source of knowledge for every task is reasonable since the focus of the empirical study is to exploit a RAG source under the assumption that it is available and of high quality, not to automatically find the best possible source of knowledge among the available ones.
For the XSS detection task, we used the XSS Evasion Cheat Sheet by OWASP\footnote{\url{https://cheatsheetseries.owasp.org/cheatsheets/XSS_Filter_Evasion_Cheat_Sheet.html}} as the RAG source.
OWASP is a foundation that works to improve the security of software and the knowledge contained in their cheatsheets is potentially useful to improve the quality of the detectors generated for XSS attack. For the SQLi detection task, the selected RAG source is an article on WebSec\footnote{\url{https://websec.wordpress.com/2010/12/04/sqli-filter-evasion-cheat-sheet-mysql/}} containing a list of SQLi patterns. The knowledge shared in both sources is well-suited to our goal of creating effective detection functions via LLMs. %
To further elaborate on the preliminary experiments conducted to select the RAG sources, we manually curated a set of sources that we considered reliable. In addition to the ones ultimately selected, for XSS detection we considered the OWASP Web Security Testing Guide\footnote{\url{https://owasp.org/www-project-web-security-testing-guide/latest/4-Web_Application_Security_Testing/07-Input_Validation_Testing/01-Testing_for_Reflected_Cross_Site_Scripting}} and a widely-read Invicti article on XSS filter evasion\footnote{\url{https://www.invicti.com/blog/web-security/xss-filter-evasion/}}. For SQL injection detection, we additionally evaluated the OWASP SQL Injection Prevention Cheat Sheet\footnote{\url{https://cheatsheetseries.owasp.org/cheatsheets/SQL_Injection_Prevention_Cheat_Sheet.html}} and the W3Schools page on SQL injection\footnote{\url{https://www.w3schools.com/sql/sql_injection.asp}}. We opted not to include these alternative sources in the main experimental configuration to avoid a combinatorial explosion in the number of experiments. Instead, we evaluated their performance by measuring the F2 scores on a representative configuration sample $(\gptf, 0.5, 2, T)$, selecting the best-performing source on the $val\_set$. For XSS detection, the two alternatives performed 6\%pt and 24\% worse than the selected source. Similarly, for SQLi detection, the OWASP Cheat Sheet and W3Schools performed 25\%pt and 33\% worse, respectively, compared to the chosen source.

\begin{table}[h]
\centering
\caption{Size of the datasets for XSS and SQLi, with a specific focus on the different class sizes and the splits into training, validation, and test set.}
\label{tab:datasets}
\resizebox{0.9\textwidth}{!}{%
\begin{tabular}{c|c|c|c|c|c|c}
\toprule
               & \multicolumn{3}{c|}{XSS}      & \multicolumn{3}{c}{SQLi}     \\
               & Training & Validation & Test & Training & Validation & Test \\
\midrule
Malicious      & 8,344     & 2,087       & 2,608 & 12,714    & 3,178       & 3,973 \\
Benign         & 8,584     & 2,146       & 2,683 & 12,714    & 3,178       & 3,973 \\
\midrule
Total & 16,928    & 4,233       & 5,291 & 25,428    & 6,356       & 7,946 \\
\bottomrule
\end{tabular}
}
\end{table}

\subsubsection{Datasets}

For XSS detection, we use a publicly available dataset containing Malicious and Benign payloads of HTTP requests from the FMereani repository\footnote{\url{https://github.com/fmereani/Cross-Site-Scripting-XSS/blob/master/XSSDataSets/Payloads.csv}}, while for SQLi detection, we use the dataset presented in the SOFIA paper~\cite{ceccato2016sofia}. Table~\ref{tab:datasets} shows the size of the splits for our two datasets, into \textit{train\_set}, \textit{val\_set}, and \textit{test\_set}. \textit{Train\_set} is not used in our approach (except for the Few-shot examples), as we generate a detection function via pre-trained LLM, without any further training. 
It is used for training the ML-based detection model, which is compared with our approach to answer to RQ3. \textit{Val\_set} is used as validation set in the TDA scenario. \textit{Test\_set} is kept hidden and it is used when answering the experimental research questions RQ1 to RQ4.

\subsection{Study Procedure}

\begin{figure}[ht]
\centering \includegraphics[width=1.0\linewidth]{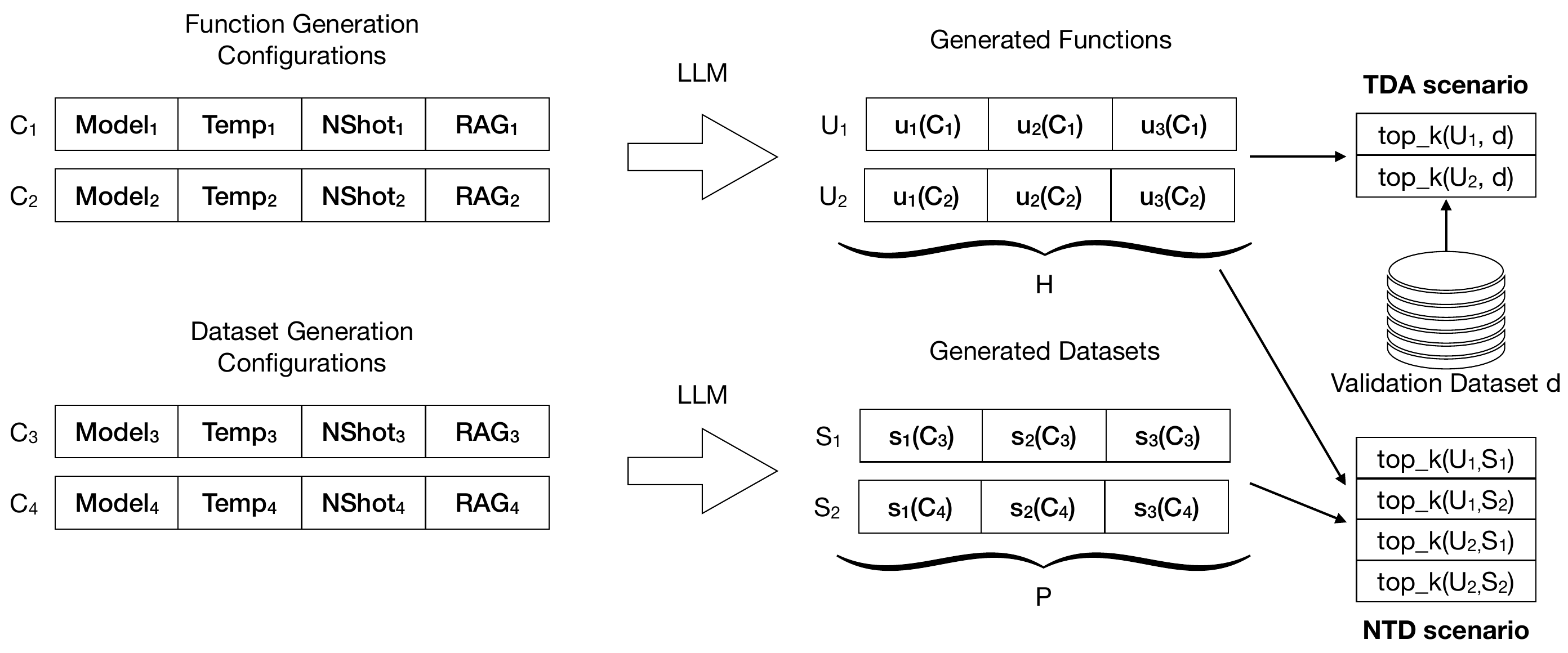}
\caption{Main steps of the adopted experimental procedure: two configurations ($C_1, C_2$) are used for function generation and two configurations ($C_3, C_4$) for dataset generation. The LLM is queried three times, resulting in three generated functions ($u_1, u_2, u_3$) and three generated datasets ($s_1, s_2, s_3$) per configuration. In the TDA scenario, the resulting functions ($U_1, U_2$) are ranked using a validation set $d$; in the NTD scenario they are ranked using the generated datasets ($S_1, S_2$).} 
\label{fig:exp-proc}
\end{figure}

The main steps of our experimental procedure are depicted in Figure~\ref{fig:exp-proc}. The pipeline at the top shows the generation of multiple functions (to account for non-determinism of the LLM) for each configuration. The pipeline at the bottom is similar, but the output consists of multiple synthetic datasets for each configuration. At the right,  Figure~\ref{fig:exp-proc} shows the selection of the best \textit{top\_k} functions based either on an existing validation dataset (top-right) or on a generated dataset (bottom-right).

\subsubsection{Function Generation} 
\label{sec:generation_output}
Given a configuration $C = \langle C.model, C.temperature, C.examples, C.rag\rangle$, and a task $t$, we denote as $u(t, C)$ the Generated Function, i.e., the output of  $C.model$ queried with the prompt constructed from the code generation template, shown in Listing~\ref{lst:code_template_rag}, the task $t$, and the parameters specified in the configuration $C$. In Figure~\ref{fig:exp-proc} (top), function generation is executed with two configurations, $C_1$ and $C_2$.

To account for non-determinism and to support Self-Ranking, we repeat the generation process $n$ (in our experiment, 40) times, obtaining a Generated Function Run $U$, consisting of the set of $n$ Generated Functions: $U(t, C) = \{u_1(t, C), \ldots, u_{n}(t, C)\}$.  In Figure~\ref{fig:exp-proc}, $n=3$ functions are generated, $u_1, u_2, u_3$ for each configuration.

 We define a Function Generation Experiment $H(t)$ as the set of all Generated Function Runs, given all valid configurations from the code generation domain (see Table~\ref{tab:confs}):
 $H(t) = \{U(t, Conf_1), U(t, Conf_2), \ldots\}$. In Figure~\ref{fig:exp-proc}, the experiment $H$ contains two sets of generated functions, $U_1$ and $U_2$.

We consider the generation of duplicates to be natural when LLM deal with code generation. Specifically, duplicates are more frequent at lower temperatures. 
We chose not to remove duplicates because they may reflect the LLM's consistency and confidence in its output. In this context, repeated solutions can be interpreted as an indication that the model assigns higher confidence to those particular outputs. 
This approach is consistent with established evaluation metrics such as pass@k \cite{chen2021evaluatinglargelanguagemodels}, which assesses the functional correctness of generated code by considering diverse solutions but does not include any mechanism to handle or account for duplicate outputs.

\subsubsection{Dataset Generation}
A synthetic dataset, denoted as $s(t, C)$, is obtained by prompting the model $C.model$ with a prompt constructed from the dataset generation template shown in Listing~\ref{lst:dataset_template}, the task $t$, and the parameters specified in the configuration $C$. %
In Figure~\ref{fig:exp-proc} (bottom), dataset generation is executed with two configurations, $C_3$ and $C_4$.

To account for non-determinism, we repeat the generation process $m$ (in our experiments, 10) times and obtain a Synthetic Dataset Run $S(t, C) = \{s_1(t, C), \ldots, s_{m}(t, C)\}$.  In Figure~\ref{fig:exp-proc}, $m=3$ datasets are generated, $s_1, s_2, s_3$ for each configuration. Given all the configurations in the domain, we define a Synthetic Dataset Experiment $P(t)$ as $\{S(t, Conf_1)$, $S(t, Conf_2)$, $\ldots\}$. In Figure~\ref{fig:exp-proc}, the experiment $P$ contains two sets of generated datasets, $S_1$ and $S_2$.

\subsubsection{Selection of \textit{top\_k} Functions}
We  select the \textit{top\_k} functions of a Generated Functions Run $U$ using a dataset $d$ and a performance metric $\mathcal{M}$ (e.g., F2-Score) by sorting the functions in $U$ by decreasing $\mathcal{M}$ and including only the first $k$ functions in the sorted list: 
$\textit{top\_k}(U, d, \mathcal{M}) = \textit{sort}(U, \mathcal{M}(\cdot, d))[1\mathpunct{:}k]$.

Note that in the event of a tie for the $k^{th}$ position, one element is randomly chosen.
In the TDA scenario, we can set $d = \textit{val\_set}$ to select the \textit{top\_k} functions. In the NTD scenario, we use a Synthetic Dataset Run as $d$, exploiting Self-Ranking.
In Figure~\ref{fig:exp-proc}, the TDA scenario is shown on the top-right, with $d$ the Validation Dataset. The NTD scenario is shown on the bottom-right, with $d$ equal to each of the generated datasets $S_1$ and $S_2$.

The usage of Self-Ranking requires black-box access to the LLM for Dataset Generation, as it is purely based on prompting, while \textit{top\_k} selection is  performed without any query to the LLM, as the ranking score is based on the generated function's performance on the synthetic dataset.

\subsubsection{Evaluation Metrics}
We use a separate, independent test set to evaluate the  results produced by the pipeline shown in Figure~\ref{fig:exp-proc}: (1) we measure the quality of the generated functions without applying any ranking (NTD scenario with no generated dataset); (2) we measure the quality of the generated functions after ranking them based on a validation set (TDA scenario); (3) we measure the quality of the generated functions after ranking them based on a generated dataset (NTD scenario with synthetic dataset).

The quality of the generated functions without ranking is measured as the average performance of the generated functions across all Function Runs $U$ in the experiment $H$. The quality of the generated functions after ranking on a validation set (resp. synthetic dataset) is measured as the \textit{top\_k} performance (i.e., the average performance of the first $k$ functions in the ranked list) of the best Generated Function Run $U^{best}$, which is the function run $U$ with highest average performance according to the validation dataset (resp. synthetic dataset). 

We also measure the \textit{top\_k} performance improvement, defined as the performance difference between the \textit{top\_k} functions and all functions in the Generated Function Run $U$.

Measuring the effectiveness of a Synthetic Dataset Run can be achieved by assessing its capability to select the \textit{top\_k} functions accurately, just as a ground-truth dataset (i.e., \textit{val\_set}) would do. 
This does not directly assess the quality of the generated Synthetic Dataset, since we cannot prove that all the elements are correctly labeled or semantically rich enough to capture all the aspects of the attack, but we indirectly assess it in terms of capability to act as a proxy for the real ground-truth dataset, thus selecting an optimal set of functions. To quantify this, we introduce the \textit{performance difference metric}, defined as the difference between the average \textit{top\_k} performance with ranking on \textit{val\_set} and the average \textit{top\_k} performance with ranking on a Synthetic Dataset Run $S$.

\subsection{Implementation}
\label{sec:impl}
Our experimental framework was implemented using Python 3.10 and Langchain\footnote{\url{https://www.langchain.com/}}. 
For comparison to SOTA, we followed the approach proposed by Chen et al.~\cite{CHEN2022102831} and implemented two XSS detection models by training a Convolutional Neural Network (CNN) and a Multi-Layer Perceptron (MLP). To compare with SOFIA~\cite{ceccato2016sofia} on SQLi detection, we consider the performance values reported in their paper, as we share exactly the same test set.
All experiments on GPT models were conducted on a machine with AMD EPYC 7742 64-Core Processor CPU, Tesla V100 GPU, 512 GB RAM, running Ubuntu 20.04.6 LTS. The experiments on other models were conducted %
on a machine with AMD EPYC 7763 64-Core Processor CPU, 32 GB RAM, running Ubuntu 22.04.4 LTS. In the latter,  models were running on services such as Google Vertex or Amazon Bedrock, depending on the provider.

\subsection{Performance Metric}
\label{sec:eval_metrics}
In our application scenarios (XSS/SQLi detection), we give more weight to false negatives (resulting in low recall) with respect to false positives (resulting in low precision), since a non-detected attack can cause much more damage than a benign request detected as an attack.
For this reason, the main performance metric used in our empirical study is the \textit{F2-Score}, referred to as \textit{F2} for simplicity ($\mathcal{M} = F_2$), which gives double importance to recall than to precision: $F_2 = 5p*r / (4p+r)$, with $p$ and $r$ indicating precision and recall, respectively.
\new{The choice of a metric that gives more weight to the recall than to the precision is related to the fact that, in many domains, an undetected attack is considered more dangerous than a false positive.}
To mitigate a possible threat to the construct validity associated with the choice of the performance metric \textit{F2}, which is strongly related to the specific domain of the study, we replicated all experiments using \textit{accuracy} and \textit{F1-Score}, referred to as \textit{F1} and defined as $F_1 = 2p*r / (p+r)$, as  alternative performance metrics.
Accuracy is a valuable choice since the datasets used in this study are well-balanced.
However, since in a real-world scenario positive samples are much fewer than the negative ones, we replicated all the experiments also using \textit{F1-Score}, a metric that ignores the false negatives and hence, differently from  accuracy, is not sensitive to the balance of positive vs negative data. Results are consistent and largely independent of the choice of the performance metric. The interested reader can find the additional results obtained with $\mathcal{M} = Accuracy$ and $\mathcal{M} = F1$ in the replication package. %

For RQ1, in the NTD scenario, we get two sets of \textit{F2} by evaluating the functions generated with and without RAG on the \textit{test\_set}.
By comparing these two sets, we check if RAG provides statistically significant improvements using the Mann-Whitney U Test~\cite{mannwhitneyu}.
To understand the impact of RAG in the TDA scenario, we use \textit{val\_set} to select the best Generated Function Run $U^{best}$ and we check if RAG was used to generate $U^{best}$.

For RQ2, we first quantify the benefits of using \textit{top\_k} selection with Self-Ranking in the NTD scenario. We analyze all pairs $\langle U, S\rangle$ of Generated Functions Runs $U$ and Synthetic Datasets Runs $S$ to compute the improvement of \textit{F2}. We employ the Wilcoxon signed-rank test to establish the statistical significance of such improvement.
To analyze the impact of Self-Ranking in the TDA scenario, we use \textit{val\_set} to select the best Function Run $U^{best}$.
Then, for each $k$, $S^{best}$ is selected starting from $U^{best}$, as the Data Set Run $S$ with minimum performance difference w.r.t. the \textit{top\_k} performance measured on \textit{val\_set}.
At this point, it is possible to measure the improvement of \textit{F2}, due to the usage of Self-Ranking, with $U^{best}$ as selected Generated Function Run and $S^{best}$ as selected Synthetic Dataset Run.

For RQ3, to compare the performance with the SOTA models, we first use \textit{val\_set} to select the best Generated Function Run $U^{best}$, then the \textit{top\_k} function of $U^{best}$. Then, we assess the \textit{top\_k} \textit{F2} measured on the \textit{test\_set}, comparing it with that of the SOTA models.

For RQ4, for each $k$, we consider one of the two tasks (e.g., XSS) in turn, and we determine the configurations that give $S_1^{best}$ and $U_1^{best}$. We then apply such configurations to produce $S_2^{transf}$ and $U_2^{transf}$ for the other task (e.g., SQLi). For the second task, we also compute $S_2^{best}$ and $U_2^{best}$.
We can now compare the best \textit{top\_k} \textit{F2} obtained using $S_2^{best}$ and $U_2^{best}$ vs. the \textit{top\_k} \textit{F2} of $S_2^{transf}$ and $U_2^{transf}$. Then, we swap the two tasks and repeat the transferability process in the other direction.

%% file: sections/results.tex
\section{Results}
\label{sec:results}

\subsection{RQ1 (RAG Benefit)}
\label{sec:rq1}

\begin{figure}[!h]
\centering
    \subfloat[XSS Detection]{%
    \includegraphics[width=0.5\linewidth]{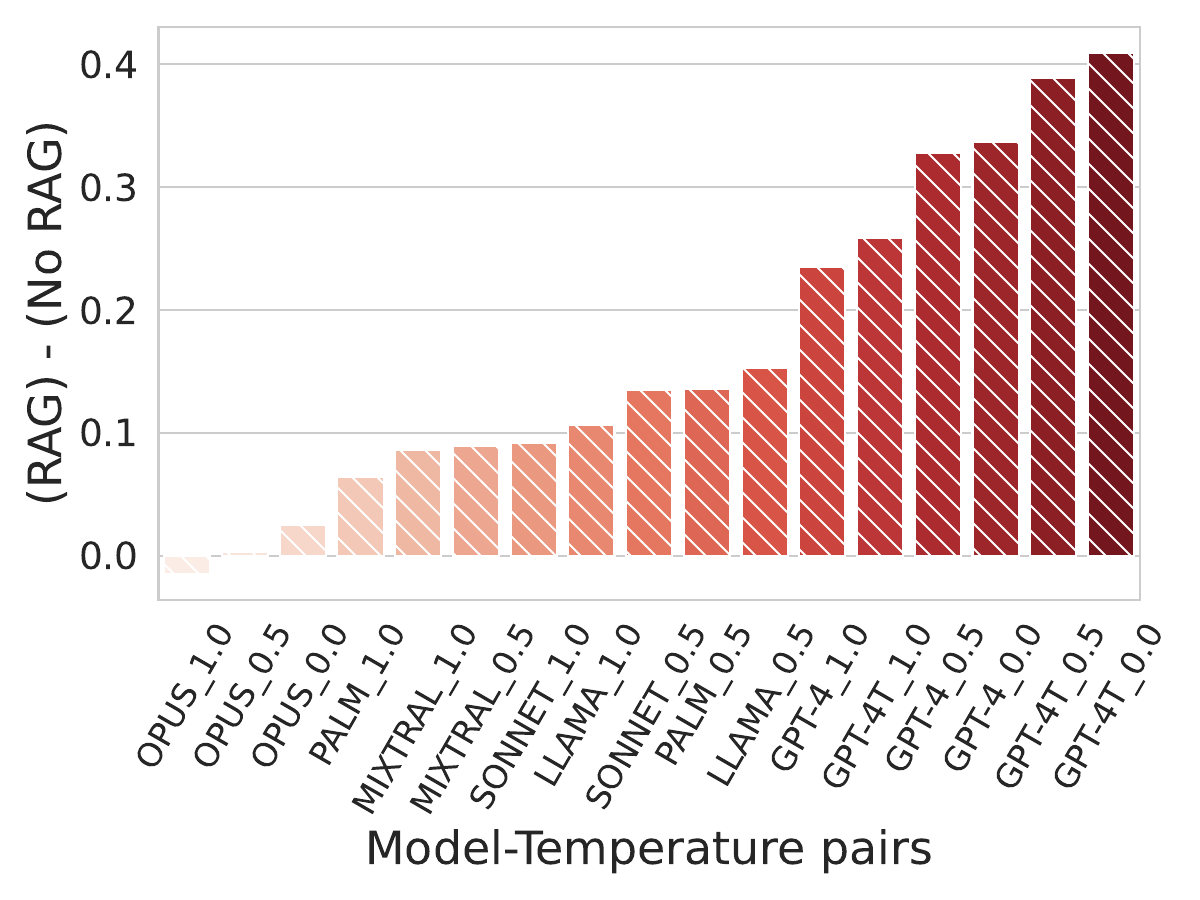}\label{fig:rq1_f2_xss}
    }
    \subfloat[SQLi Detection]{%
    \includegraphics[width=0.5\linewidth]{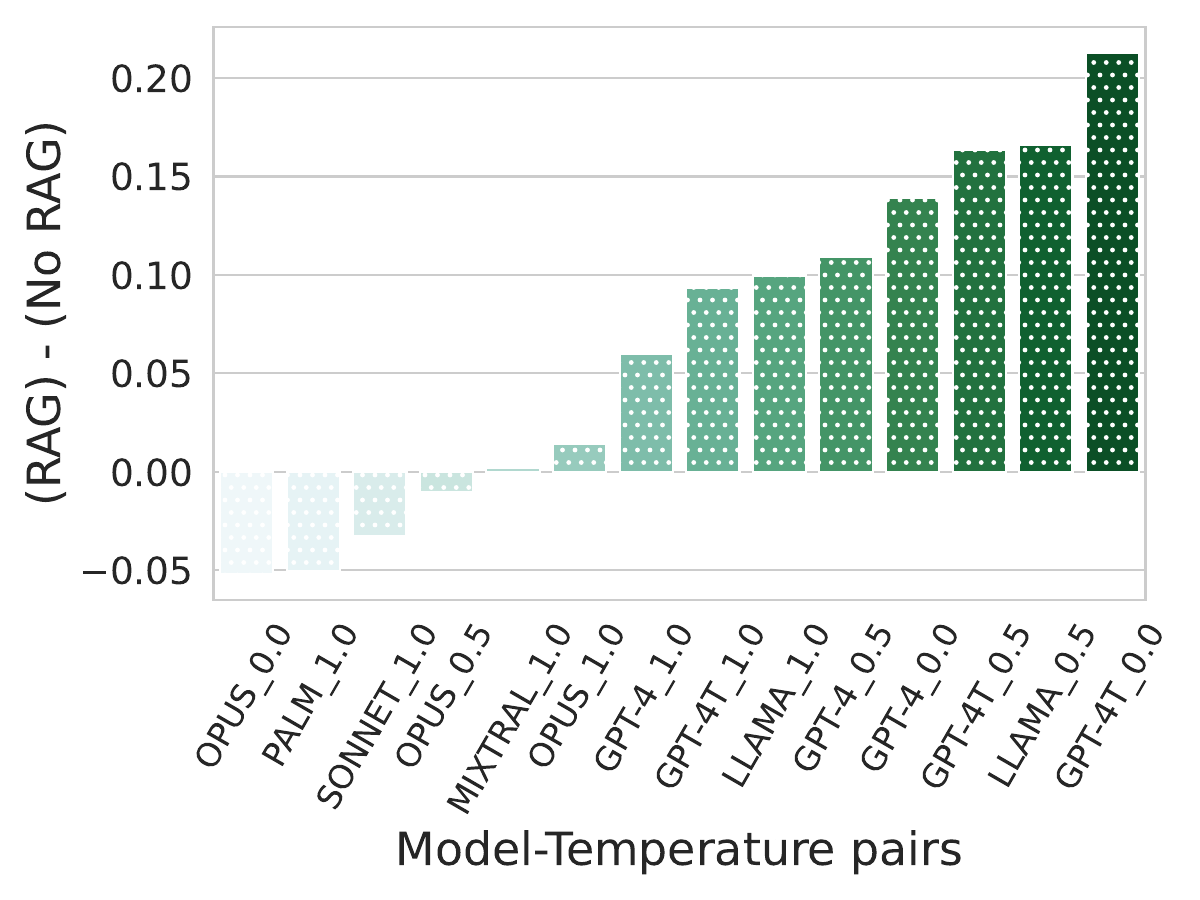}\label{fig:rq1_f2_sqli}
    }
\caption{Difference between the F2 of Generated Function Runs \textbf{with RAG} (i.e., RAG Prompt and RAG-Few-shot Prompt) and the F2 of Generated Function Runs \textbf{without RAG} (i.e., Basic Prompt and Few-shot Prompt), for XSS detection (left) and SQLi detection (right).}\label{fig:rq1}
\end{figure}

Figure~\ref{fig:rq1} illustrates the impact of RAG on function generation for XSS and SQLi in the NTD scenario, showing the F2 differences between the configurations with and without RAG, across all possible configurations. Each bar represents the average F2 score difference for a specific Model-Temperature pair when using RAG, compared to the same pair without RAG. Results indicate that employing RAG generally enhances the performance of function generation for both tasks. The number of Model-Temperature pairs benefiting from RAG is much larger than the number of pairs showing degradation, and the improvements are statistically significant, as evidenced by $p$-values ($\approx 10^{-66}$ for XSS and $\approx 10^{-24}$ for SQLi) below the standard threshold of 0.05. 
\new{Furthermore, models such as GPT-4, which demonstrated superior performance without RAG, also exhibit greater improvements when RAG is employed. This suggests that the baseline knowledge of these models is not only more robust but their capacity to utilize the contextual information provided by RAG is also more effective.}

\begin{figure}[!h]
\centering
    \subfloat[XSS Detection]{%
    \includegraphics[width=0.5\linewidth]{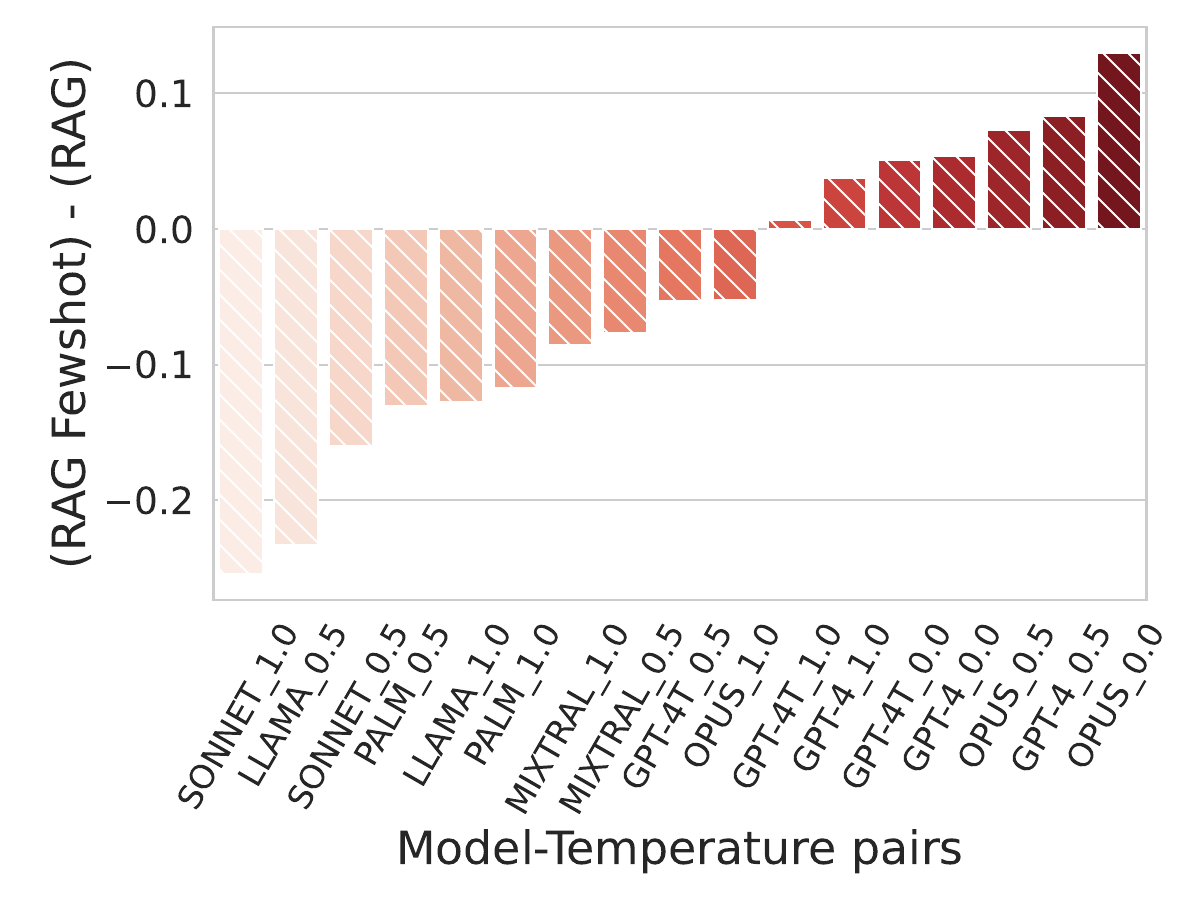}\label{fig:rag_fs_rag_xss_f2}
    }
    \subfloat[SQLi Detection]{%
    \includegraphics[width=0.5\linewidth]{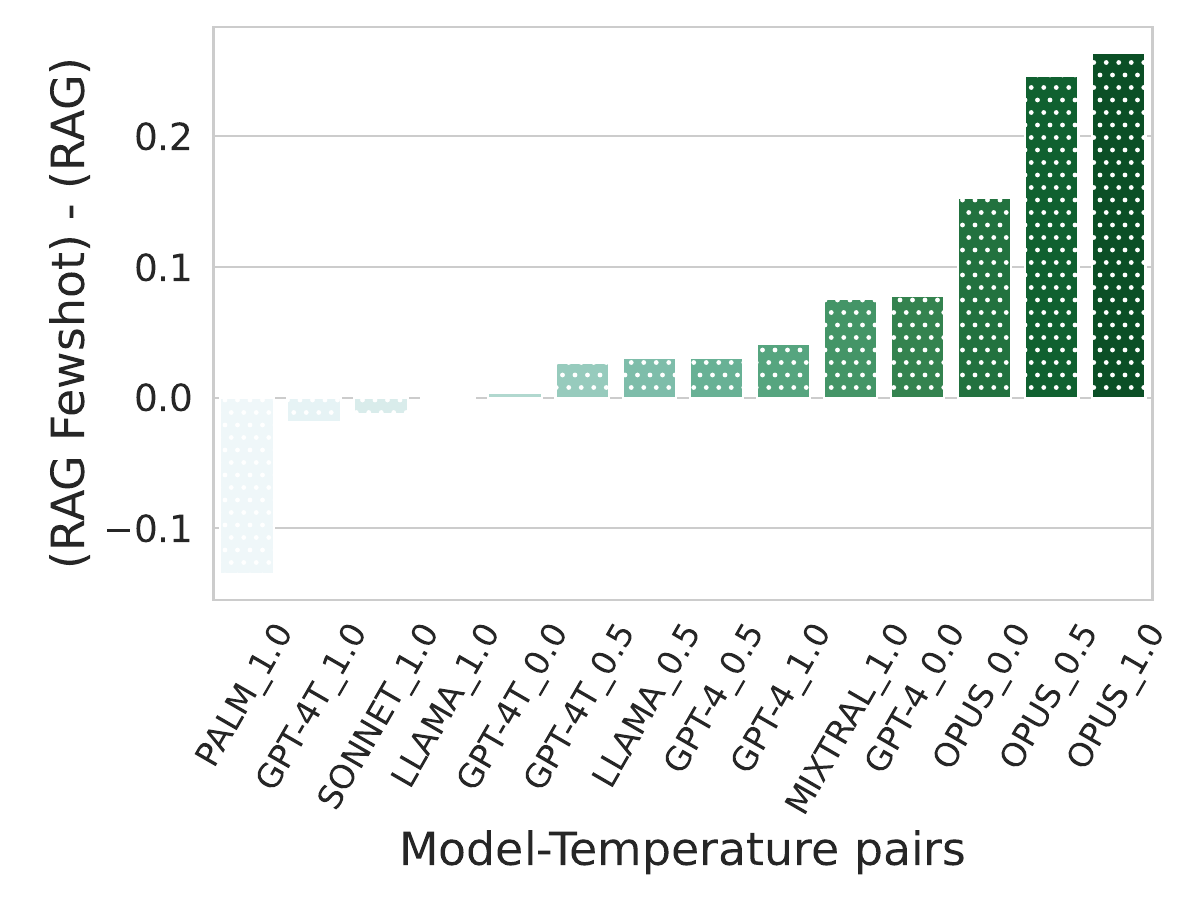}\label{fig:rag_fs_rag_sqli_f2}
    }
\caption{Difference between the F2 of Generated Function Runs \textbf{with RAG Few-shot Prompt} and the F2 of Generated Function Runs \textbf{with RAG Zero-shot Prompt}, for XSS detection (left) and SQLi detection (right).}\label{fig:rq1_few_shot}
\end{figure}

We further investigate the benefit of combining Few-shot examples with RAG using a similar setting as Figure~\ref{fig:rq1}. Figure~\ref{fig:rq1_few_shot} shows that
while the addition of Few-shot examples shows some benefits for SQLi, the same cannot be said for XSS. These findings suggest that the usage of Few-shot examples may not always provide advantages when RAG is already employed, indicating that in the NTD scenario, it could be preferable to omit them.

\begin{figure}[!h]
\centering
    \subfloat[XSS Detection]{%
    \includegraphics[width=0.5\linewidth]{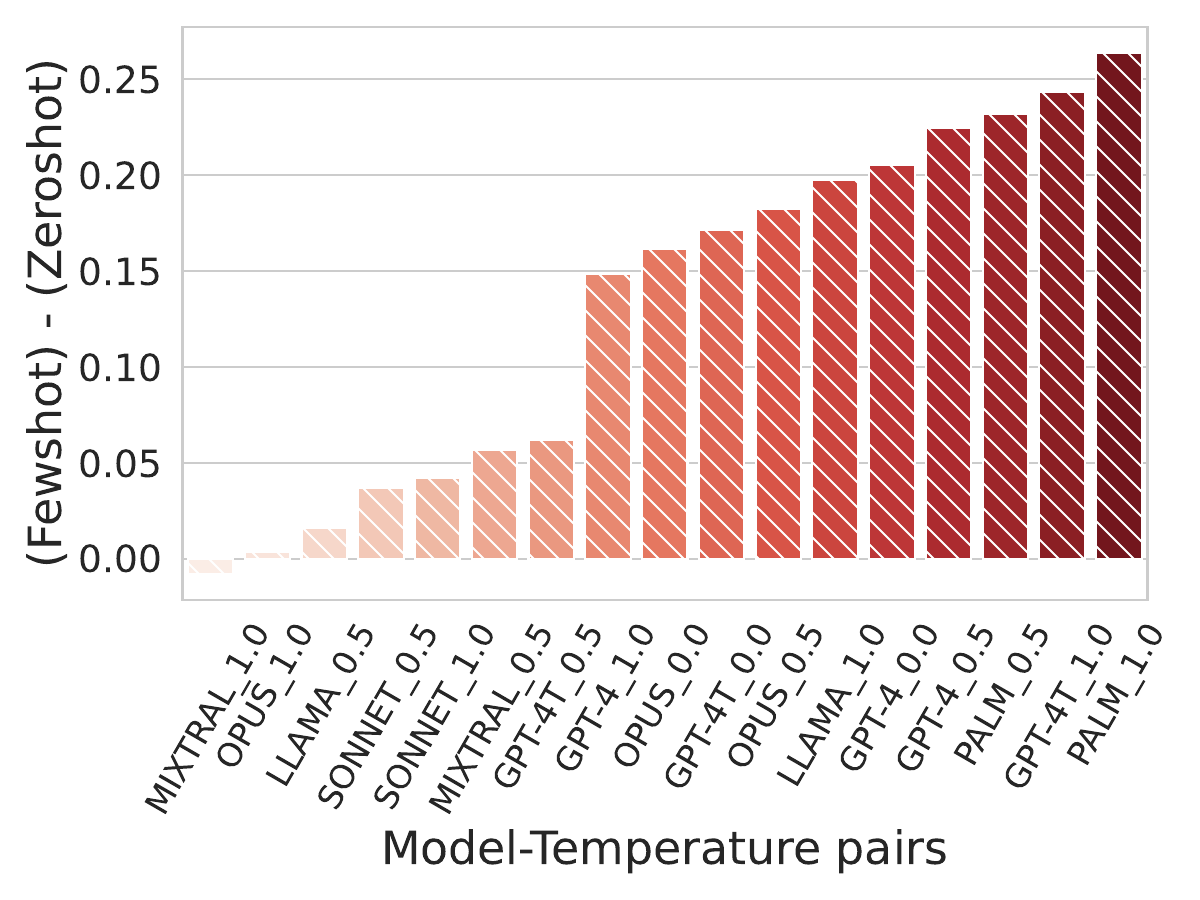}\label{fig:fs_zero_xss}
    }
    \subfloat[SQLi Detection]{%
    \includegraphics[width=0.5\linewidth]{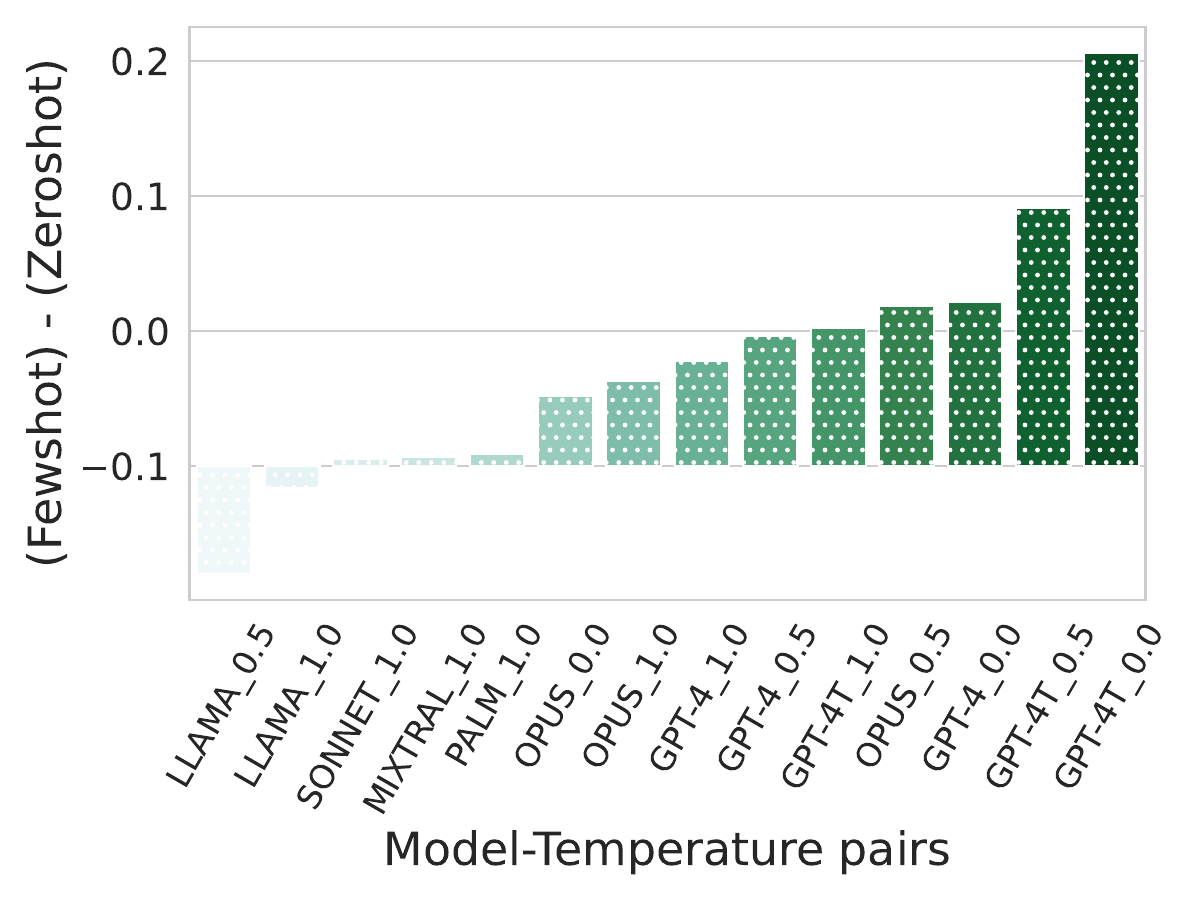}\label{fig:fs_zero}
    }
\caption{Difference between the F2 of Generated Function Runs \textbf{with Few-shot Prompt} and the F2 of Generated Function Runs \textbf{with Zero-shot Prompt}, for XSS detection (left) and SQLi detection (right).}\label{fig:rq1_few_shot}
\end{figure}

However, as illustrated in Figure~\ref{fig:rq1_few_shot}, Few-shot examples offer a significant advantage when RAG is not employed. This is likely due to the well-documented challenge LLMs face in effectively retrieving and utilizing specific knowledge from large contexts~\cite{liu-etal-2024-lost}. In the absence of RAG, Few-shot examples play a crucial role in guiding the model. However, when RAG is used, the extensive amount of provided knowledge appears to overshadow the influence of the Few-shot examples, rendering them less effective within the broader context.

Turning to the TDA scenario where we can select the best Generated Function Run using $val\_set$, we observe that $U^{best}$ takes always advantage of  RAG as well as Few-shot examples: for XSS, $U^{best}=$ (\gptft, 0.0, 10, T), and for SQLi, $U^{best}=$ (\gptft, 0.0, 2, T). For brevity, we omit the full results table for the TDA scenario here, but it is included in the replication package. This indicates that while the use of few-shot examples may not consistently enhance performance in general (as seen in the NTD scenario), the best configuration selected with $val\_set$ (in the TDA scenario) leverages the benefits of using both RAG and Few-shot examples.

\begin{tcolorbox}[boxrule=0pt,frame hidden,sharp corners,enhanced,borderline north={1pt}{0pt}{black},borderline south={1pt}{0pt}{black},boxsep=2pt,left=2pt,right=2pt,top=2.5pt,bottom=2pt]
\textbf{Answer to RQ1}:
The usage of RAG yields statistically significant performance improvements for both tasks. While the combination of RAG with Few-shot examples does not consistently enhance this benefit, it proves advantageous in the best case.
\end{tcolorbox}

\subsection{RQ2 (Impact of Self-Ranking)}
\label{sec:rq2}
\begin{figure}[h]
\centering
    \subfloat{%
    \includegraphics[width=0.33\linewidth]{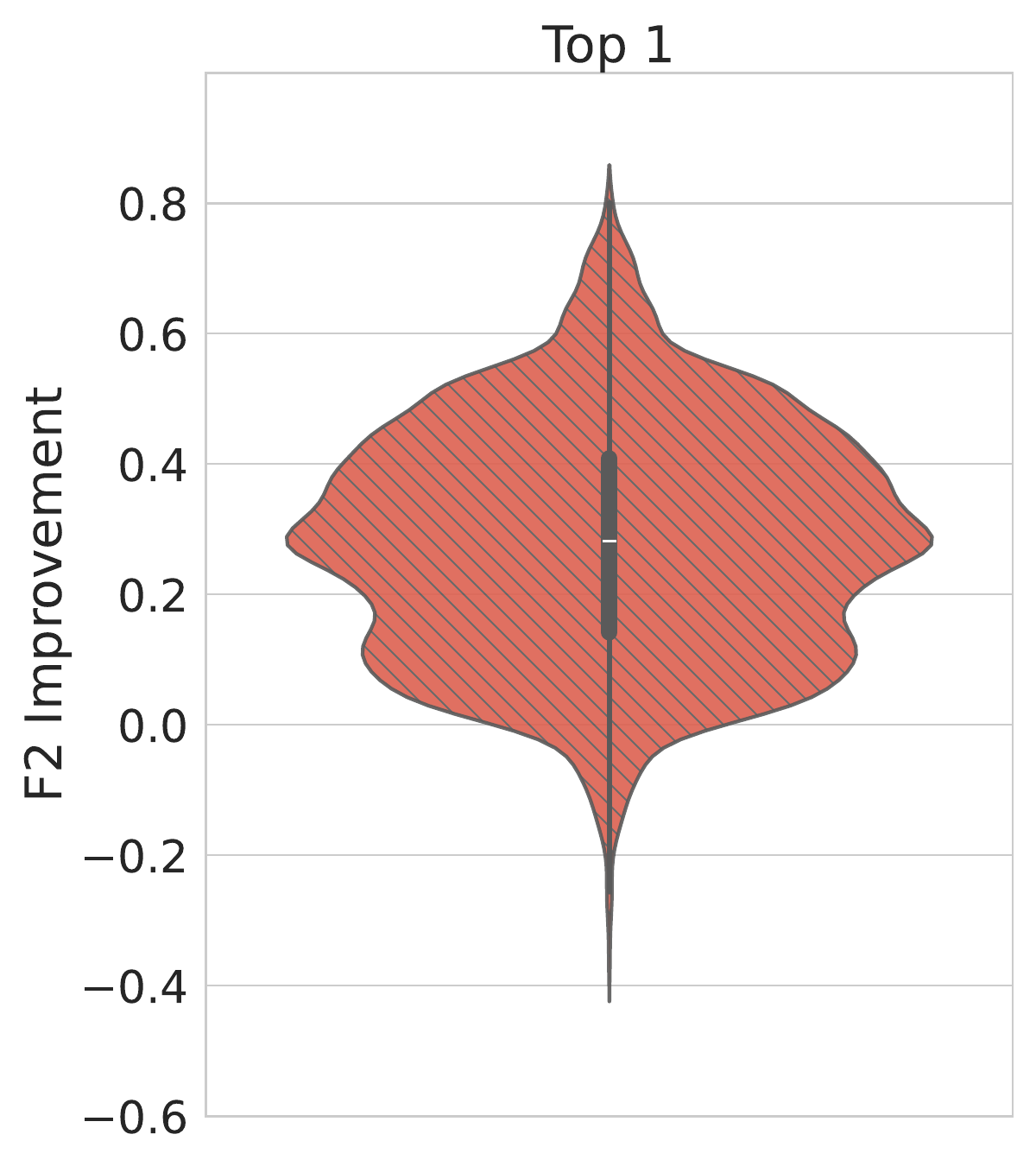}\label{fig:rq2_top_1_f2_xss}
    }
    \subfloat{%
    \includegraphics[width=0.33\linewidth]{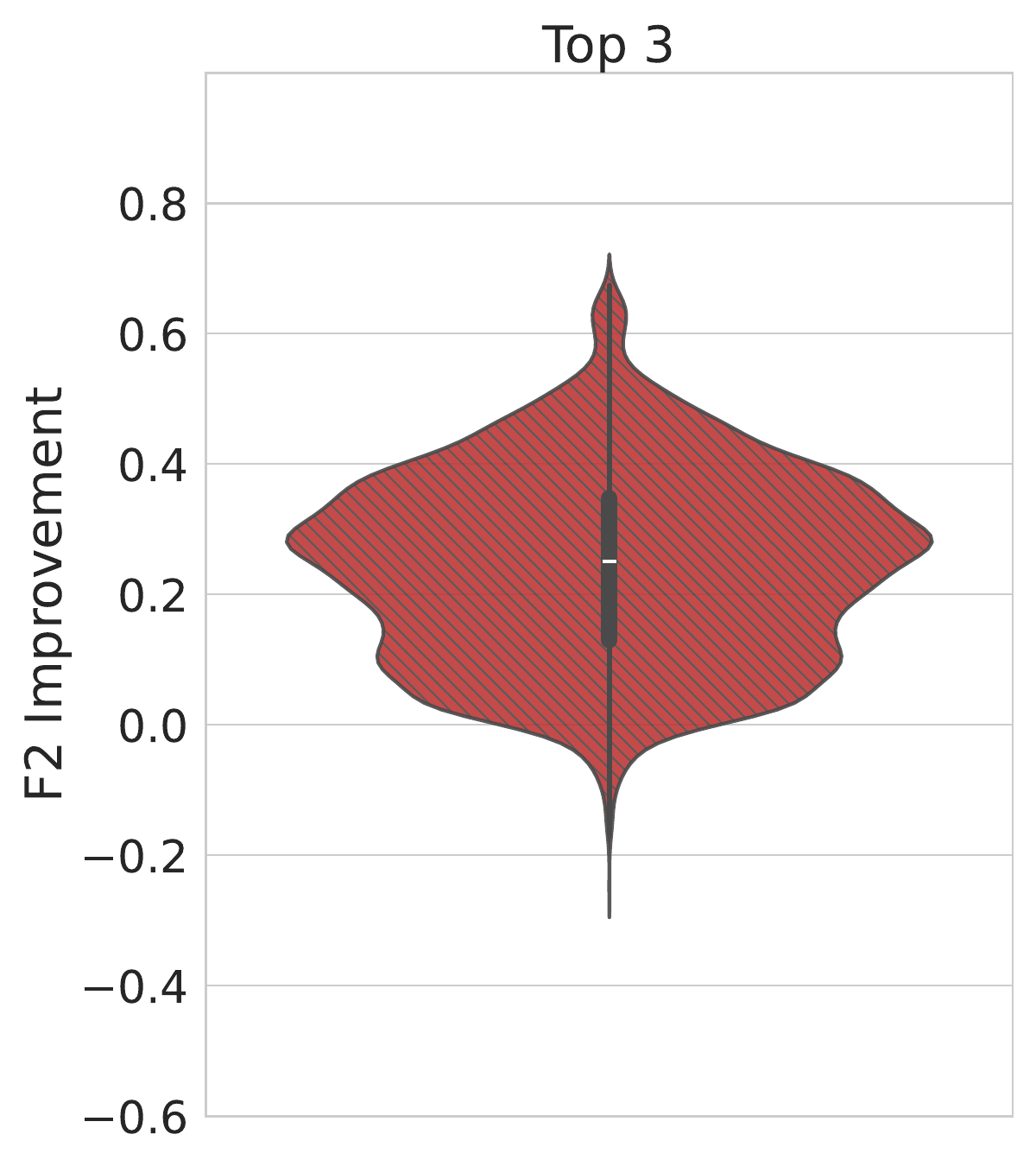}\label{fig:rq2_top_3_f2_xss}
    }
    \subfloat{%
    \includegraphics[width=0.33\linewidth]{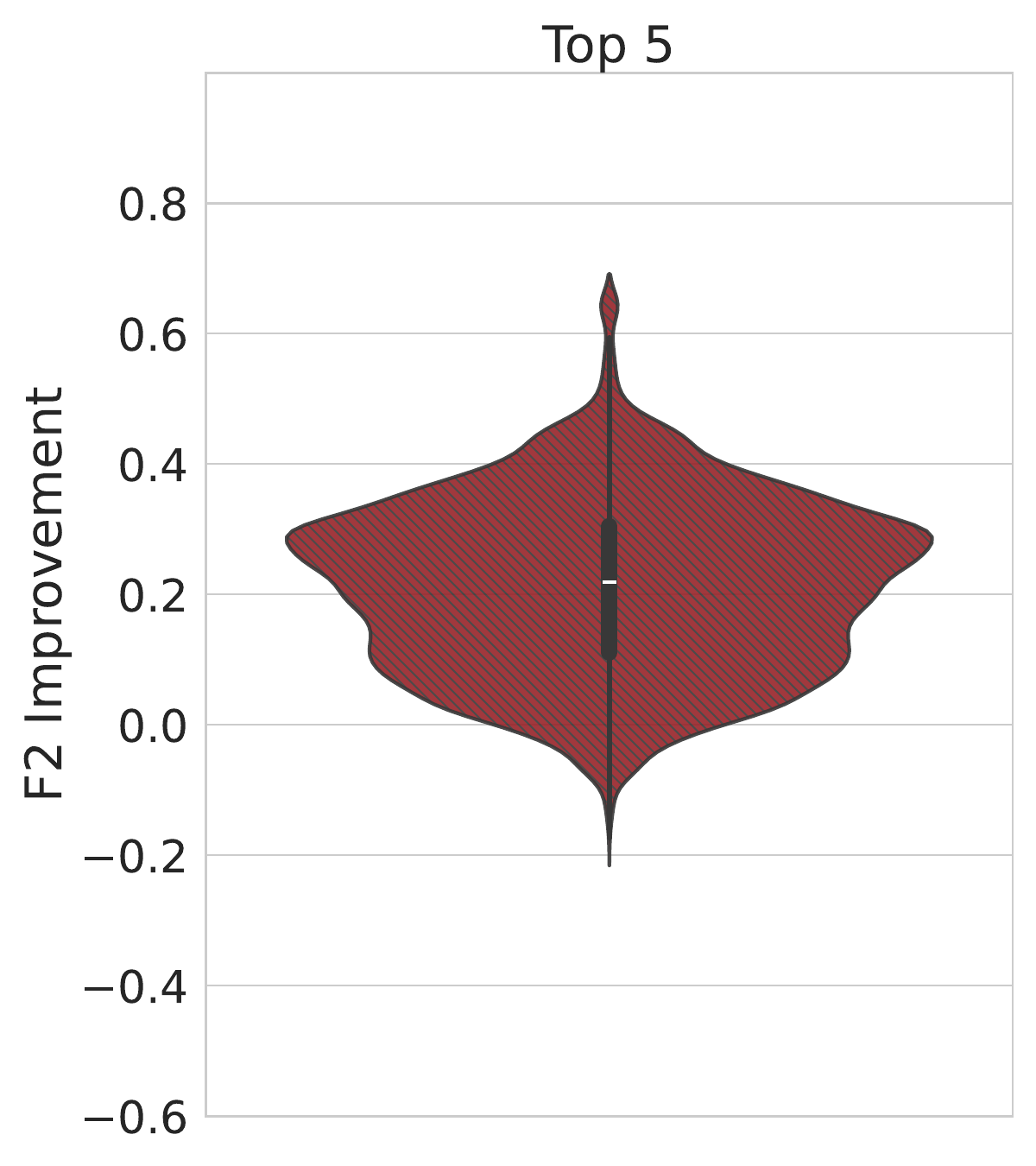}\label{fig:rq2_top_5_f2_xss}}\\
    
    \parbox{\linewidth}{\centering XSS Detection\\[1ex]}

    \subfloat{%
    \includegraphics[width=0.33\linewidth]{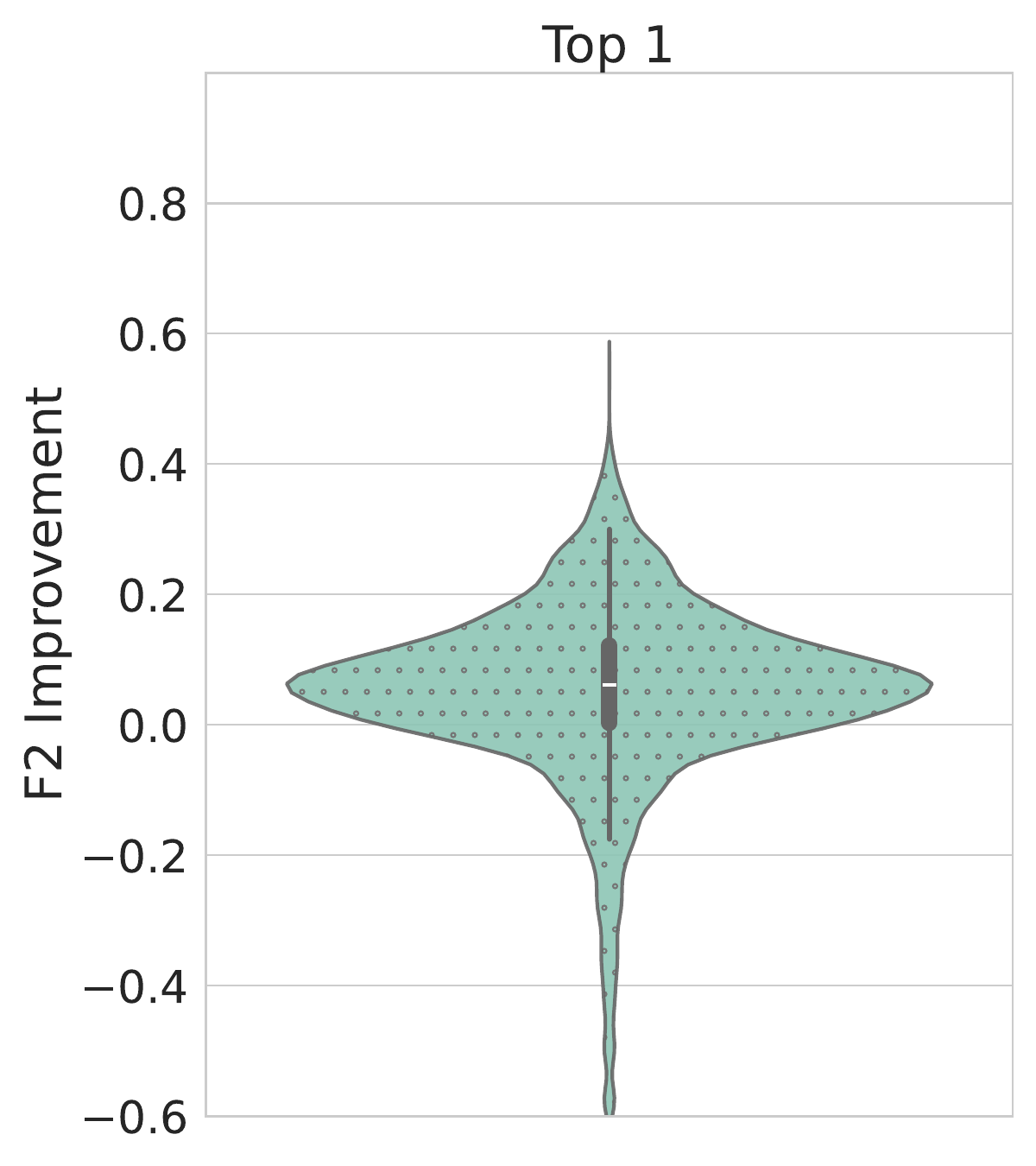}\label{fig:rq2_top_1_f2_sqli}
    }
    \subfloat{%
    \includegraphics[width=0.33\linewidth]{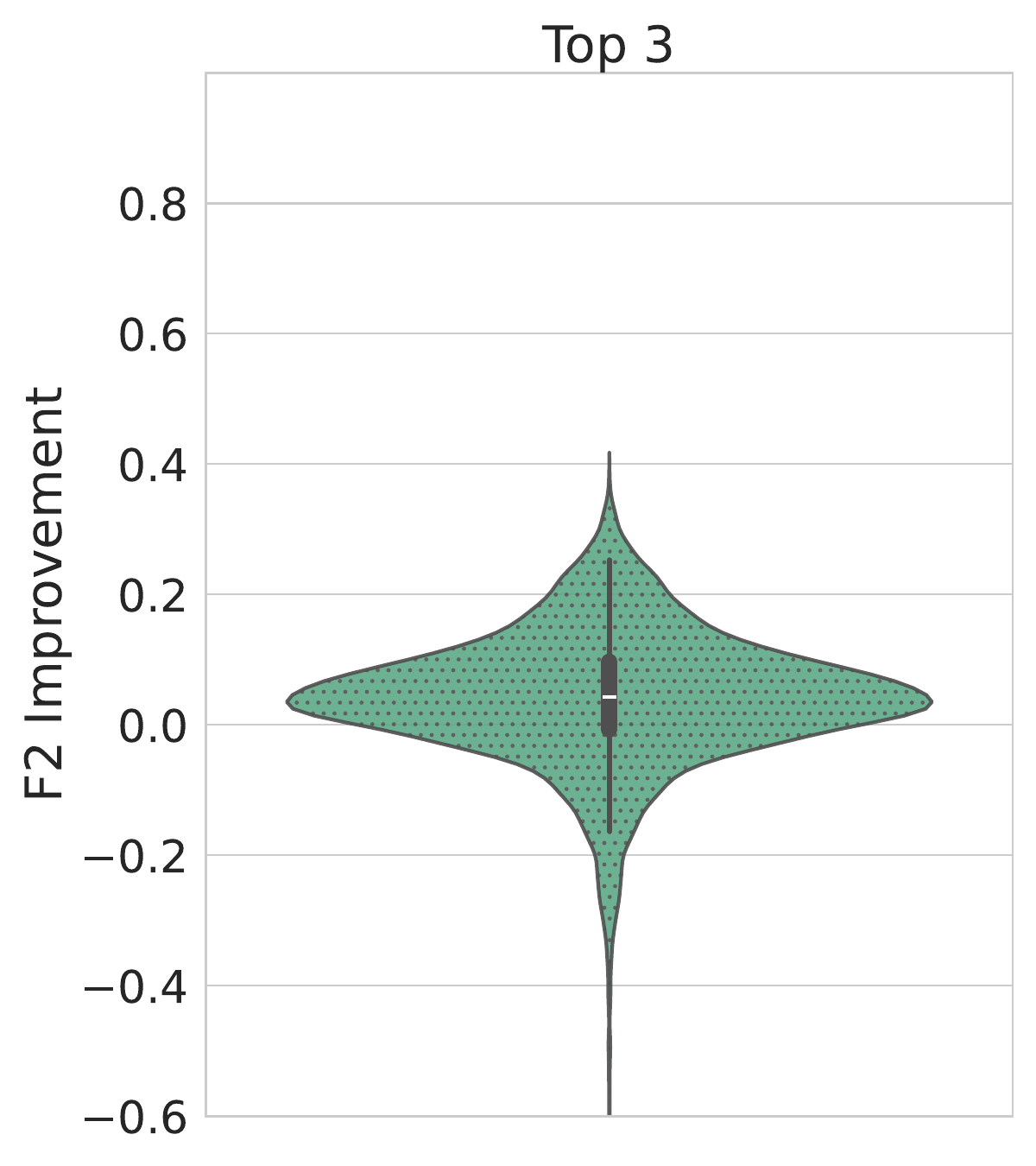}\label{fig:rq2_top_3_f2_sqli}
    }
    \subfloat{%
    \includegraphics[width=0.33\linewidth]{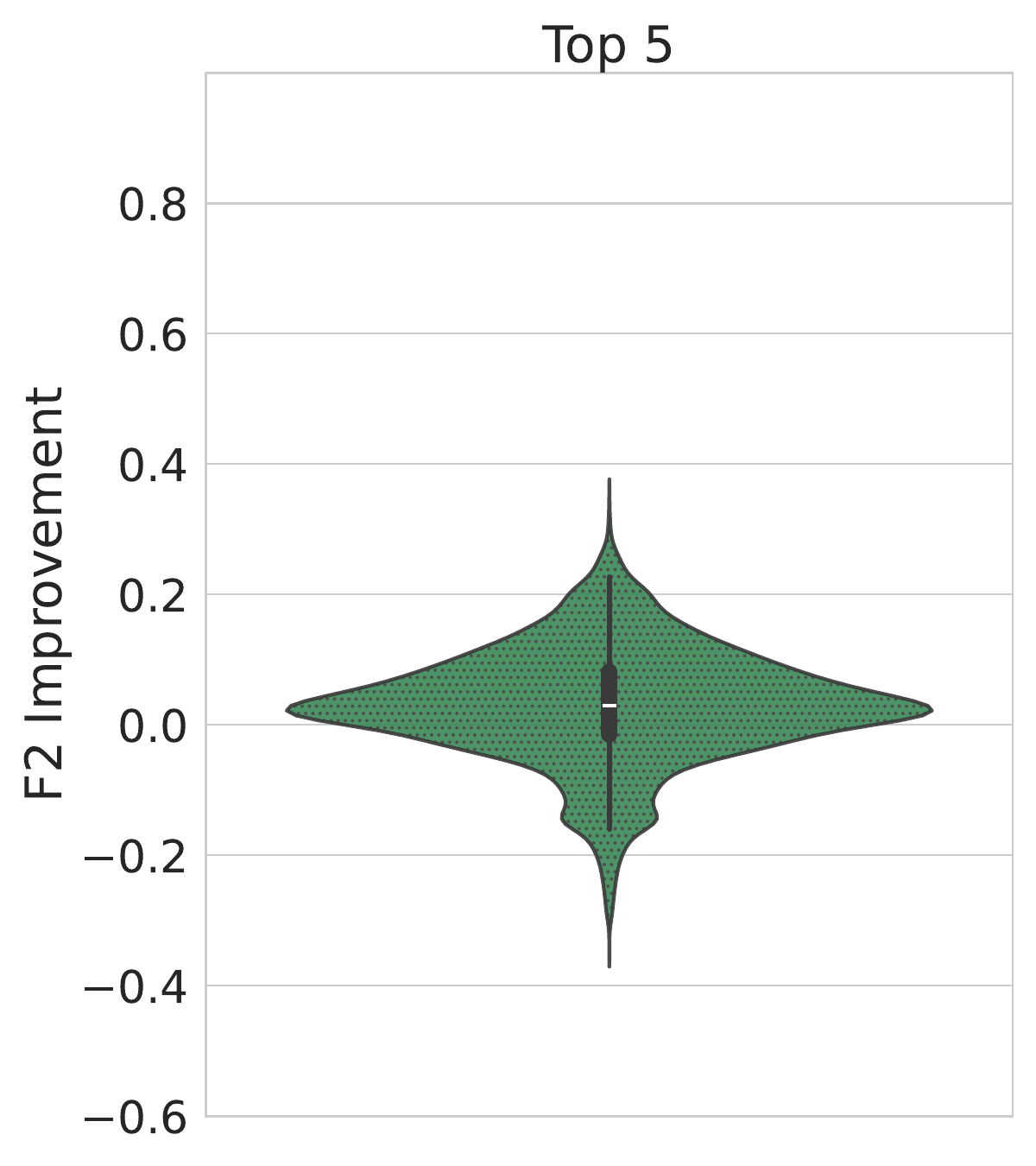}\label{fig:rq2_top_5_f2_sqli}
    }\\
        \parbox{\linewidth}{\centering SQLi Detection\\[1ex]}

\caption{F2 given by \textit{top\_k} selection (i.e., Self-Ranking) for XSS detection (first row) and SQLi detection (second row), with $k=1$, $k=3$ and $k=5$.}\label{fig:rq2}
\end{figure}

Let us consider the NTD scenario.
The violin plots shown in Figure~\ref{fig:rq2} depict the effect of Self-Ranking, i.e., $top\_k$ selection across three values of $k$ for the two tasks, considering all the possible pairs of function/synthetic dataset configurations. Overall, we observe a clear improvement, particularly for XSS. The region between quartiles for XSS falls largely between a 20\%pt and 40\%pt improvement, with an average improvement of 37\%pt (up to 71\%pt) and an improvement that affects 98\% of the cases. While the improvement for SQLi is less pronounced, we still observe improvements in 73\% cases, with an average improvement of 6\%pt (up to 43\%pt). Additionally, we can see that, as $k$ increases, the average improvement decreases for both tasks, but the gains become more stable.
The improvements given by the usage of $top\_k$ selection are statistically significant: the $p$-values obtained with the Wilcoxon signed-rank test are below the threshold of 0.05 for both tasks and all the values of $k$\footnote{Specifically, $p$-value is $0$ for all the $k$s for XSS. For SQLi, it was $0$ when $k=1$ and $k=2$ and $\approx 10^{-247}$ when $k=5$.}.

In the TDA scenario, we observe that utilizing the Self-Ranking mechanism to select the \textit{top\_k} functions is more effective than not employing it: when compared to solely using $U_{best}$, employing $S_{best}$ achieves 3.21\%pt and 4.94\%pt increases in F2, for XSS and SQLi respectively.

The Synthetic Datasets generated with the LLMs proved to be effective in improving the final performance, exploiting Self-Ranking. However, for completeness, it is also interesting to directly assess the quality of the generated Synthetic Datasets. To have a better overview of the quality of those datasets, we randomly sampled 100 Malicious and Benign samples from various Synthetic Datasets and we asked  three human evaluators with many-year expertise in the security domain to label them as Malicious or Benign. 

Given an example of Malicious/Benign HTTP/SQLi payload generated inside a Synthetic Dataset, the task for the evaluators was to label it as Malicious or Benign in a simple multiple-choice question. The question was preceded by a description of the task and a few ground-truth examples (both for XSS and SQLi). An example of a Benign HTTP Payload is \textit{http://www.github.com/user/repo}; \textit{http://example.com/?search=alert('XSS')} is an example of a Malicious HTTP payload, while \textit{UPDATE users SET password = '1' or '1' = '1', is\_admin = 1} and \textit{SELECT * FROM users WHERE id = 1} are two examples of, respectively, a Malicious and a Benign SQL query.

After selecting the annotation as the majority vote among the three annotators, the agreement with the label assigned by the LLM while generating the samples was 98\%. This result underlines that the generated Synthetic Datasets are not just a good tool to be exploited by Self-Ranking, but they are a good-quality artifact that can be potentially exploited in a larger set of tasks.

\begin{tcolorbox}[boxrule=0pt,frame hidden,sharp corners,enhanced,borderline north={1pt}{0pt}{black},borderline south={1pt}{0pt}{black},boxsep=2pt,left=2pt,right=2pt,top=2.5pt,bottom=2pt]
\textbf{Answer to RQ2}:
Utilizing the Self-Ranking mechanism that selects $top\_k$ functions leads to statistically significant improvements on both tasks.
\end{tcolorbox}

\begin{table}[h]
\caption{Comparison with SOTA models.}
\label{tab:rq3}
\centering
\begin{tabular}{c|r|c|r}
\toprule
\multicolumn{2}{c|}{\textbf{XSS Detection}} & \multicolumn{2}{c}{\textbf{SQLi Detection}}                                  \\
\midrule

Method  & \multicolumn{1}{c|}{F2} & Method  & \multicolumn{1}{c}{F2} \\
\midrule
CNN~\cite{CHEN2022102831}             &   0.998    & \multirow{2}{*}{SOFIA~\cite{ceccato2016sofia}} & \multirow{2}{*}{0.993} \\

MLP~\cite{CHEN2022102831}                &    0.995   &                                               &                   \\
\midrule
Ours (k=1)        &   0.965   & Ours (k=1)                             &    0.991               \\
Ours (k=3)        &  0.965     & Ours (k=3)                             &   0.988                 \\
Ours (k=5)        &  0.965     & Ours (k=5)                             &   0.975                \\
\midrule
Baseline (Few-shot)    &   0.922   & Baseline (Few-shot)                        &    0.882 \\
Baseline (Basic Prompt)      &   0.630   & Baseline (Basic Prompt)                           &    0.800  \\
\bottomrule
\end{tabular}%
\end{table}

\subsection{RQ3 (Comparison with SOTA)}
\label{sec:rq3}

To compare ours with learning-based SOTA techniques, we select $U^{best}$ and $top\_k$ functions based on $val\_set$. For XSS, $U^{best}$ is $(\gptft, 0.0, 10, T)$, while for SQLi, it is $(\gptft, 0.0, 2, T)$.
Due to lack of space, we omit the table with all the results, which is anyway available in the replication package. We establish two baselines to better understand the improvement offered by adopting our approach. The first baseline is obtained using Basic Prompt, without Few-shot examples and RAG.
It represents the expected results that can be obtained by generating detectors without exploiting any of the techniques presented in this work.
The second one is obtained by considering also the usage of Few-shot examples, and it represents the expected results that can be obtained without exploiting any of the techniques that are the direct contributions of this work.

Table~\ref{tab:rq3} presents the results of the comparison. We observe a significant improvement of 34\%pt and 18\%pt for XSS and SQLi respectively, when compared to the Basic Prompt baseline. 
The improvement with respect to the Few-shot Baseline is of 4.3\%pt and 10\%pt for XSS and SQLi respectively.
There is a slight decrease in F2 compared to SOTA models, with a 3.15\%pt and 0.83\%pt drop for XSS and SQLi, respectively. 
We argue that the slight performance gap between our approach and SOTA models is understandable, given our approach's training-free nature and direct applicability to multiple tasks. Moreover, these results provide empirical support for our claim that incorporating external knowledge and Self-Ranking is essential for LLMs to achieve competitive performance with SOTA models. 

It is important to remark that, despite the slight performance decrease w.r.t. SOTA models, there are several other advantages associated with the use of our method: 
our approach is applicable even in the absence of a training set, a scenario that rules out all ML-based SOTA techniques. We use a pre-trained LLM, while SOTA techniques require that a model is designed and trained for the attack detection task at hand. The output of our approach (a Python function) is interpretable by developers, while SOTA models are black-box.
This transparency allows for a clear and full understanding of the detector's decision-making process and provides an opportunity for further refinement and improvement.
Another big advantage is  transferability from one detection task to another, an aspect we will explore in the following section, which is structurally impossible for  SOTA models.

\begin{tcolorbox}[boxrule=0pt,frame hidden,sharp corners,enhanced,borderline north={1pt}{0pt}{black},borderline south={1pt}{0pt}{black},boxsep=2pt,left=2pt,right=2pt,top=2.5pt,bottom=2pt]
\textbf{Answer to RQ3}:
The functions generated with our approach are shown to be comparable to the SOTA models trained for the specific tasks, with the advantage of interpretability and applicability in the absence of a training set.
\end{tcolorbox}

\begin{table}[h]
\caption{Best Generated Function Runs and Synthetic Datasets}
\label{tab:rq4_runs}
\centering
\begin{tabular}{c|c|c}
\toprule
\multicolumn{3}{c}{\textbf{Task 1: XSS Detection}} \\
\midrule
$U^{best}_1 \rightarrow U^{transf}_2$ &$k$ &
  $S^{best}_1 \rightarrow  S^{transf}_2$ 
  \\
  \midrule
\multirow{3}{*}{(\gptft, 0.0, 10, T)}
  &
  1 & (\haiku, 0.5, 6, F)  \\ 
  
  & 3 &(\haiku, 1.0, 10, F)  \\ 
  & 5 & (\haiku, 0.5, 10, F)  \\
 \midrule 
 
 \multicolumn{3}{c}{\textbf{Task 2: SQLi Detection}} \\
\midrule
$U^{best}_2 \rightarrow  U^{transf}_1$ &$k$ &
  $S^{best}_2 \rightarrow  S^{transf}_1$ 
   \\
  \midrule
  \multirow{3}{*}{(\gptft, 0.0, 2, T)} 
   & 1 &(\opus, 0.5, 6, T)  \\
   & 3 & (\opus, 0.5, 6, T)  \\
  & 5 &   (\sonnet, 1.0, 10, T)   \\
  \bottomrule
\end{tabular}%
\end{table}

\begin{table}[h]
\caption{Results of transferability study. The results with transferred configurations are underlined.}
\label{tab:rq4}
\centering
\begin{tabular}{c|c|c|c}
\toprule
\multicolumn{4}{c}{\textbf{Task 1: XSS Detection}} \\
\midrule
\multicolumn{1}{c|}{$k$}&
  \multicolumn{1}{c|}{$F2(U^{best}_1, S^{best}_1)$} &
 \multicolumn{1}{c|}{Avg. F2} &
 \multicolumn{1}{c}{$F2(U^{transf}_1, S^{transf}_1)$} \\
  \midrule
1 & 0.965 & 0.809 & \underline{0.949} \\
3 & 0.965 & 0.771 & \underline{0.929} \\
5 & 0.965 & 0.740 & \underline{0.933} \\
\midrule
\multicolumn{4}{c}{\textbf{Task 2: SQLi Detection}} \\
\midrule
 \multicolumn{1}{c|}{$k$}&
 \multicolumn{1}{c|}{$F2(U^{best}_2, S^{best}_2)$} &
 \multicolumn{1}{c|}{Avg. F2} &
 \multicolumn{1}{c}{$F2(U^{transf}_2, S^{transf}_2)$} \\
  \midrule
1 & 0.964 &
  0.787 &
  \underline{0.853}  \\ 
  3 &  0.951 & 
  0.775 &
  \underline{0.900}  \\
  5 & 0.946 &
  0.763 &
  \underline{0.867}  \\
  \bottomrule
\end{tabular}%
\end{table}

\subsection{RQ4 (Transferability)}
\label{sec:rq4}

Table~\ref{tab:rq4_runs} shows the best configuration of each task, specifically for function generation ($U^{best}$) and synthetic dataset generation ($S^{best}$), across different $k$ values. In our notation, Task 1 is XSS and Task 2 is SQLi, and $A \rightarrow B$ represents the transfer of a configuration from one task to another. As an illustration, in $U^{best}_1 \rightarrow U^{transf}_2$, $U^{best}_1$ denotes the best configuration for Task 1, whereas $U^{transf}_2$ represents the \textit{transferred} configuration, which originates from Task 1 ($U^{best}_1$) and is subsequently  evaluated on Task 2.

Table~\ref{tab:rq4} presents the transferability results. The `$F2(U^{best}, S^{best})$' columns indicate F2 computed on the original task with its best configuration, `Avg. F2' columns represent the average F2 computed across all the $U-S$ pairs for a given $k$, and the `$F2(U^{transf}, S^{transf})$' columns show F2 computed using \textit{transferred} configurations. Comparing the transferred configuration's performance to the average column provides a good estimate of the benefits of configuration transfer over a mere random selection of a configuration for a new, unseen task.

The results support the effectiveness of transferring configurations. While there is a slight degradation in F2 compared to the original, best configuration (on average, 3\%pt for XSS and 8\%pt for SQLi), we can observe that results of transferred configurations outperform the average F2, achieving on average, 16\%pt improvement for XSS and 10\%pt improvement for SQLi.

\begin{tcolorbox}[boxrule=0pt,frame hidden,sharp corners,enhanced,borderline north={1pt}{0pt}{black},borderline south={1pt}{0pt}{black},boxsep=2pt,left=2pt,right=2pt,top=2.5pt,bottom=2pt]
\textbf{Answer to RQ4}:
The best configurations obtained for one task can effectively be transferred to the other task, outperforming the average performance across all possible configurations.
\end{tcolorbox}

%% file: sections/discussion.tex
\begin{figure}[!h]
\centering
    \subfloat{%
    \includegraphics[width=0.5\linewidth]{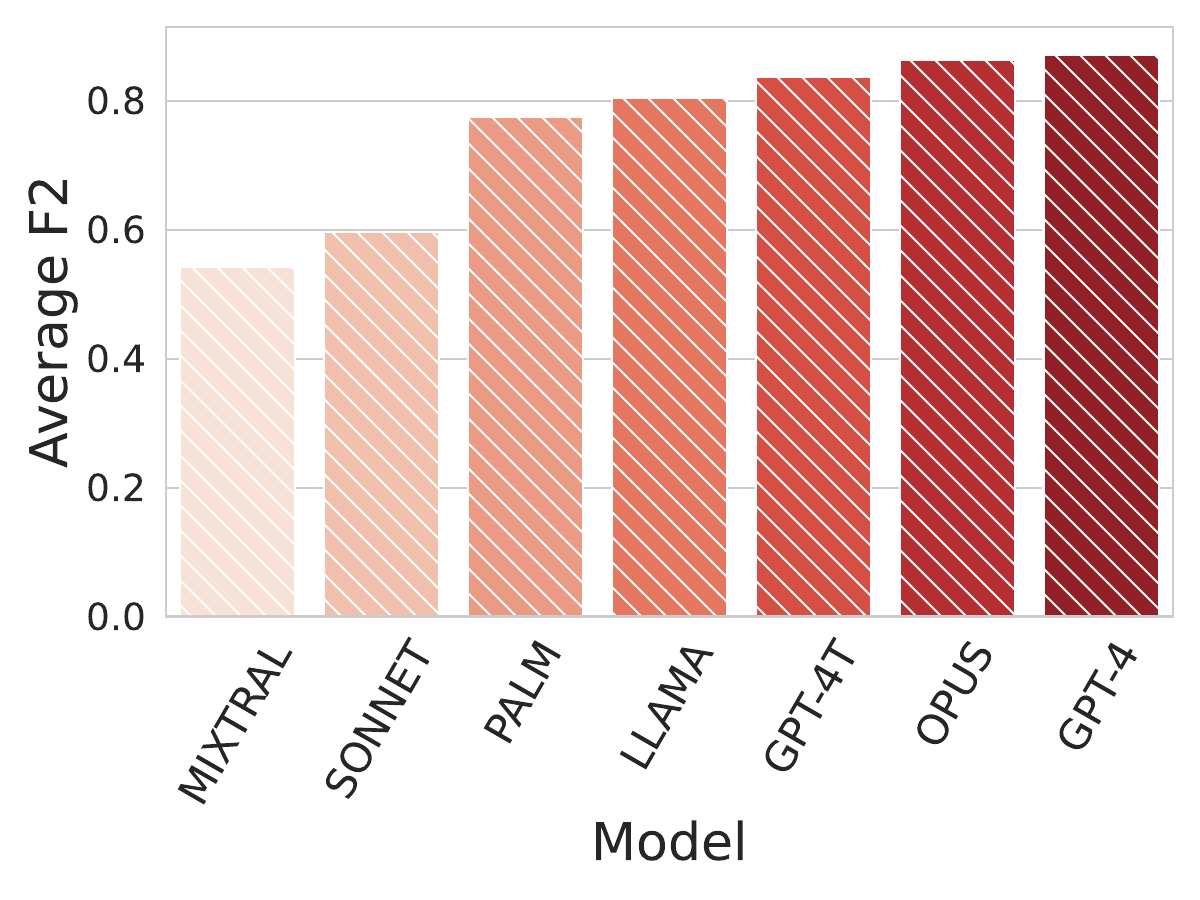}\label{fig:xss_f2}
    }
    \subfloat{%
    \includegraphics[width=0.5\linewidth]{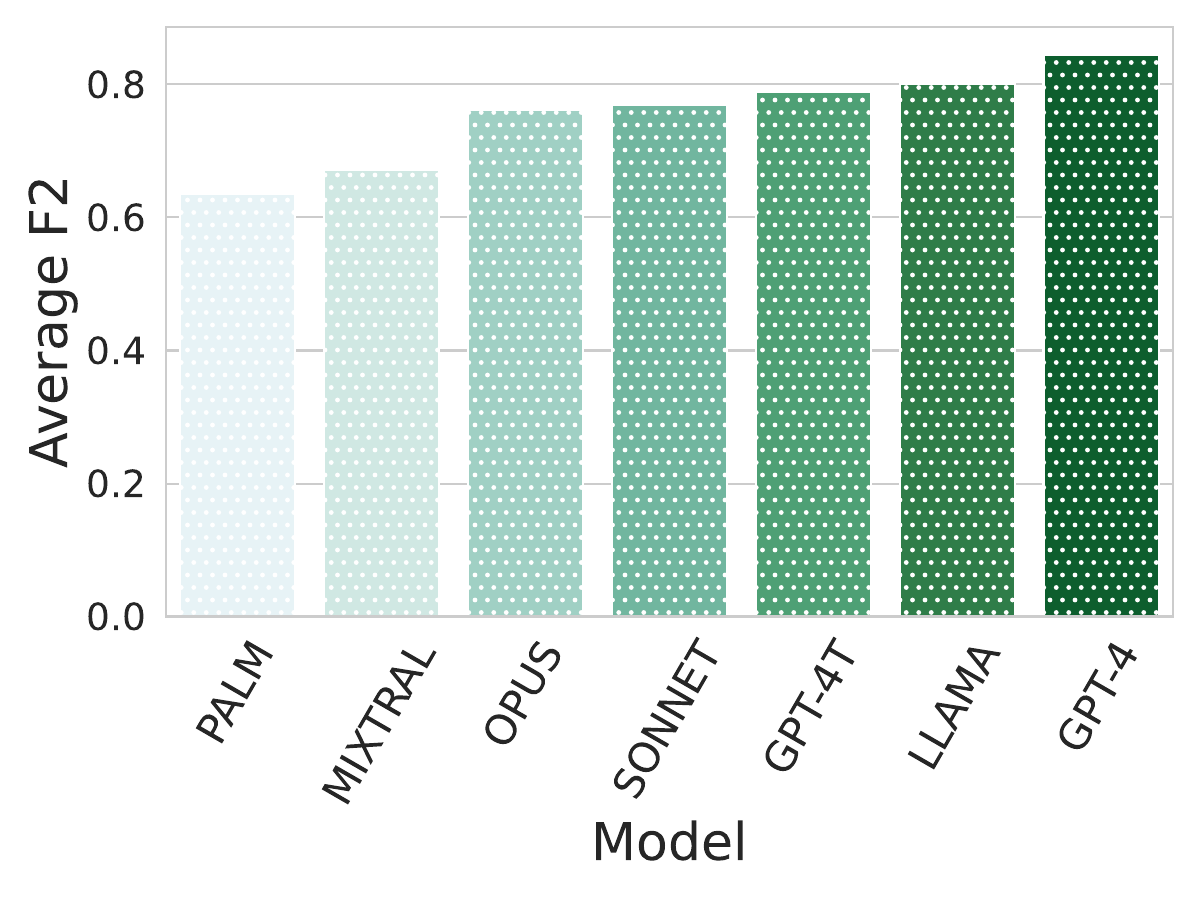}\label{fig:sqli_f2}
    }
\caption{Average F2, broken down by each LLM, for XSS detection (left) and SQLi detection (right).}\label{fig:f2}
\end{figure}

\section{Discussion}
\label{sec:discussion}

With RQs 1 and 2, we highlighted the benefits of using RAG and Self-Ranking, highlighting the necessity to incorporate external knowledge about the attacks to generate robust detectors, and exploiting the multiple reasoning-paths of the LLM to select the $K$ top-ranked detectors. However, these results do not offer explicit insights into the best choices for the other parameters, such as LLM or temperature. While this question may have a trivial answer in the TDA scenario where developers can experiment and evaluate different configurations, the NTD scenario presents a more challenging situation. On one hand, our transferability study (RQ4) demonstrated that the best-performing combinations for one task  exhibit strong performance on the other task. Therefore, the combinations from another task could serve as a reasonable starting point for selecting the inner parameters when approaching a new task. 
Since we considered just two tasks,  results are not enough to show generalizability to  tasks different from the considered ones. Still, they suggest potential transferability under similar task characteristics.
When transferability is not possible (e.g., because developers cannot access the LLM identified via transferability), developers have still to choose a model and a temperature. We can support their choice by conditioning the results of our experiments on model or on temperature. Results conditioned for each model are shown in Figure~\ref{fig:f2}, which presents the average F2 achieved by different LLMs across the two tasks. These results indicate the strong performance of \gptft and \gptf, which is unsurprising given their inclusion within the $U^{best}$ in the earlier experiments. On the other hand, \mixtral consistently underperforms relative to other LLMs in both tasks. While these plots exhibit some correlation with the HumanEval scores reported in Table~\ref{tab:models}, it is important to note that models like \haiku, which achieved decent HumanEval scores, were discarded due to their inability to successfully complete the task (see Section~\ref{sec:exp_settings}). This observation aligns with our hypothesis that while HumanEval effectively assesses the reasoning capabilities of LLMs, it may not directly translate to the depth of knowledge required for more specialized tasks, such as secure code generation.

Regarding the temperature, our analysis suggests that it is highly LLM-dependent, making it challenging to draw general conclusions across LLMs. One consistent finding relates to synthetic dataset generation, where higher temperatures tend to outperform lower temperatures. This is because lower temperatures often result in a lack of diversity in generated examples, leading to a limited variety of samples.

Regarding the knowledge sources used in RAG, we selected them manually. However, assuming the deployment of the proposed method in a real-world setting, we can envision an automated retrieval process that leverages a predefined set of reliable sources, such as those from OWASP, to extract relevant information without the need for manual curation.

\new{Our approach assumes that a developer aiming to use the LLM generation process can identify and access a suitable RAG source for their target task. They must also be able to retrieve or create a few representative examples to guide the generation process and either craft an effective task-specific prompt from scratch or adapt one of our provided prompts (e.g., those designed for XSS or SQLi) to suit their specific use case.}

\new{When considering the choice between our LLM-based approach and SOTA trainable models like DNN, MLP, or CNN, the decision depends on the availability of a labeled dataset. Our results demonstrate that, in the absence of a large dataset, our methodology can achieve results comparable to SOTA without this prerequisite.}

%% file: sections/related_work.tex
\section{Related Work}
\label{sec:related_work}

Recent research has scrutinized the security of code generated by LLMs, showing that it contains vulnerabilities. Mousavi et al.~\cite{mousavi2024investigation} conducted an examination of Java code produced by ChatGPT in the context of security API use cases. They uncovered 20 distinct varieties of API misuses by ChatGPT, demonstrating insecure practices. Bhatt et al.~\cite{bhatt2024cyberseceval} presented a dataset exposing insecure code generation by LLMs, encompassing 9 programming languages and addressing 50 prevalent vulnerabilities. Khoury et al.~\cite{khoury2023secure} instructed ChatGPT to generate 21 programs susceptible to vulnerabilities including XSS and SQLi. While these works covered a broader range of vulnerabilities than our work, their focus remained primarily on \textit{identifying} insecure code generation of LLMs. In contrast, our research aims to systematically \textit{enhance} the security of the LLM-generated code. We achieve this by employing RAG and Self-Ranking. We also evaluated the LLM-generated functions systematically, using large datasets available from SOTA techniques. This contrasts with prior works that used only a few test cases to obtain evidence of vulnerabilities in LLM-generated security attack detectors.

Nair et al.~\cite{nair2023hwsecurecode} investigated ChatGPT-induced vulnerabilities in hardware code. They explored multiple prompts and provided guidance on generating secure code. While their goal aligns with ours, their evaluation was not conducted systematically using existing, solid benchmarks. Moreover, their approach to enhancing the security of generated code relies on manually crafted adjustments to the prompt, tailored to individual vulnerabilities. This manual intervention constrains the scalability and adaptability of their approach, particularly when addressing a broad spectrum of vulnerabilities.

SVEN~\cite{he2023large} is a learning-based approach that aims to guide LLM's code generation towards satisfying a given property, exploiting Prefix-Tuning~\cite{li2021prefix} to fine-tune the LLM. While SVEN demonstrated promising results in preventing LLMs from introducing vulnerabilities, it requires a fine-tuning procedure based on a curated training set, hindering its efficiency and applicability to closed LLMs, hence limiting its scope.%
To the best of our knowledge, our work is the first work that generates
attack detectors for SQLi and XSS instead of training a DNN to solve these tasks or focusing on vulnerability detection in existing code.

%% file: sections/threats.tex
\section{Threats to Validity}
\label{sec:threats}

\textbf{Internal validity}. 
LLMs are non-deterministic, hence we repeated our function and synthetic dataset generation experiments 40 and 10 times respectively (the disparity is due to the higher stability of the latter experiments). Indeed, 
we adopt Self-Ranking to exploit the non-deterministic nature of LLMs. While different LLM configurations and selection of RAG sources may yield varying results, our choices were based on documentation and existing best practices.

\textbf{External validity}. Our approach used nine recent LLMs and they were evaluated on two prevalent vulnerabilities, XSS and SQLi. While our approach may not generalize to all other vulnerabilities, we believe that our approach of enhancing prompts with external knowledge and Self-Ranking is a robust and general method. %

Our approach was evaluated on two tasks where high-quality documentation--used as RAG sources in our experiments--was available. However, for other tasks, obtaining knowledge sources of comparable quality might be challenging. In such cases, the overall performance of our method may change. Specifically, the positive impact of RAG is likely to be reduced, while Self-Ranking retains its potential to contribute.

Since we are using publicly available datasets, data contamination is a possible concern in this work, as in  any LLM-based work. However, since the functions generated by the LLM alone are  far from perfect, it is reasonable to conjecture that data contamination is not a significant issue in our case.

\textbf{Construct validity}. We utilized F2-Score and Accuracy as our evaluation metrics, which are considered standard measures in the security domain.

\textbf{Conclusion validity}. Many elements in our approach are non-deterministic. For this reason, we draw our conclusions based on statistical non-parametric tests (Mann-Whitney U and Wilcoxon).

%% file: sections/conclusion.tex
\section{Conclusion}
\label{sec:conclusion}

In this paper, we present a novel approach to improving the robustness of LLM-generated security attack detectors by integrating RAG and Self-Ranking into the prompting process. Our extensive study with nine LLMs targets two well-known and prevalent vulnerabilities, Cross-Site Scripting (XSS) and SQL injection (SQLi). Results show the effectiveness of our approach in improving the robustness of the detectors generated by the LLM. The integration of external knowledge with RAG resulted in a notable enhancement in detection performance while Self-Ranking further improved the results. Our findings provide valuable insights for developers, highlighting the importance of incorporating relevant knowledge and utilizing automated methods for the assessment of LLM-generated code.
At the same time, these results open the way for the researchers to develop optimal strategies to perform RAG using a snippet of code as the query and documentation as a source of knowledge, since this strategy can potentially benefit those application domains where knowledge-intense code generation is prevalent.